\documentclass[aps,prb,floatfix,twocolumn,showpacs,preprintnumbers]{revtex4-1}
\usepackage{amsmath,amssymb,graphicx}
\usepackage{filecontents}
\usepackage[utf8]{inputenc}
\usepackage[T1]{fontenc}
\usepackage{xcolor}
\usepackage{float}
\usepackage{array}
\newcolumntype{C}[1]{>{\centering\arraybackslash}m{#1}}
\usepackage{booktabs}
\bibpunct{[}{]}{;}{n}{}{}
\usepackage[para,online,flushleft]{threeparttable}
\usepackage{multirow}

\usepackage{caption}
\usepackage{subfig}

\usepackage{ragged2e}
\DeclareCaptionJustification{justified}{\justifying}
\IfFileExists{newtxtext.sty}
   {\usepackage{newtxtext,newtxmath}}
   {\IfFileExists{stix.sty}
      {\usepackage{stix}}
      {\IfFileExists{mathptmx.sty}
      {\usepackage{mathptmx}}{} } }

\usepackage{textcomp}

\usepackage{bm}
\usepackage{bbold}
\renewcommand{\vec}[1]{\mathbf{#1}}

\newcommand*\chem[1]{\ensuremath{\mathrm{#1}}}

\IfFileExists{siunitx.sty}{\usepackage{booktabs,siunitx}}{}

\pdfoutput=1
\usepackage{color}
\definecolor{LinkColor}{rgb}{0.256,0.439,0.588}
\usepackage{hyperref}
\hypersetup{
   pdfauthor={Huu T. Do and K. S. D. Beach},
   pdftitle={Effective interactions between local hopping modulations on the square lattices},
   colorlinks=true,
   citecolor=LinkColor,
   linkcolor=LinkColor,
   urlcolor=LinkColor,
}

\usepackage{textcomp,marvosym}

\usepackage{braket}

\usepackage{physics}

\begin{document}

\title{Effective interactions between local hopping modulations on the square lattice}

\author{Huu T. Do}
\email{htdo@go.olemiss.edu}
\author{Khagendra Adhikari}
\author{K. S. D. Beach}
\email{kbeach@olemiss.edu} 
\affiliation{Department of Physics and Astronomy, The University of Mississippi, University, Mississippi 38677, USA}

\date{\today}

\begin{abstract}
We address the problem of free fermions interacting with frozen gauge fields. In particular, we consider a tight-binding model of fermions on the square lattice in which (i) flux 0 or $\pi$ is threaded through each plaquette and (ii) each nearest-neighbor link is decorated with an Ising degree of freedom that describes the local modulation of the hopping amplitude. Following the standard Ruderman--Kittel--Kasuya--Yosida (RKKY) approach, we compute an effective spin model in the coupling strength order by order. Unlike the original RRKY result for site-centered SU(2) spins in which the leading contribution is an effective exchange term at the second-order, perturbation theory in link-centered Z$_2$ case produces a first-order term that favors a collective ferromagnetic (FM) moment. If, by some means, an antiferromagnetic (AFM) configuration can be stabilized, the energetics of ground state is controlled by an effective Ising interaction acting pairwise at the long range across the system.

\end{abstract}
\maketitle
\section{Introduction} 
	 
	Novel properties of two dimensional (2D) lattices such as the integer quantum Hall effect in graphene \citep{Kane} and superconducting phenomenon in the \chem{CuO_2} plane (high-$T_\text{c}$ superconductors) are attractive to investigate in detail. Recently, the observation of practical single-layer ferromagnet with the finite Curie temperature $T_\text{C}$ can be applied for fabricating spintronic and magneto-electric devices \citep{Huang}. The tight-binding models in the square lattice with adding some generic interacting terms are used to describe the emergent properties of the superconducting order in cuprate- and iron-based superconductors \citep{Schattner}. 

	Observing gapless Fermi surface in the cuprate compounds by ``resonating valence bond'' theory was the primary reason that Affleck \textit{et al}.\ proposed the $\pi$-flux model on the square lattice \citep{Affleck}. By applying perpendicularly the constant magnetic field through the lattice, the hopping integral between fermions at the adjacent sites is coupled with the gauge-invariant magnetic flux \citep{Harris}. When the total flux passing through each plaquette is exact $\pi$, the gauge-inequivalent state is formed, and the unit cell tunes from single basis into bipartite one. The Fermi surface with finite volume shrinks into four Dirac points in the energy dispersion \citep{Otsuka, Xu}. 
	
	The models of 0-flux \citep{Schattner} and the $\pi$-flux \citep{Assaad2, Xu, Gazit} square lattices reemerging in the context of coupling fermions with the transverse field Ising spins show various exotic ground-state phases and quantum phase transitions. In those models, an Ising spin variable on the link modifies the hopping amplitude of fermions at the neighboring sites. The first example of Ising-nematic quantum phase in the square lattice has been  proposed to understand the formation of phase diagram of the Fe-based superconductors \citep{Schattner}. Second, Assaad \textit{et al}.\ has presented the existence of different ground-state phases and quantum phase transitions by changing the number of fermions per site \citep{Assaad2}. The rich variety of quantum phase transitions such as the first-order transition of two different topological ferromagnetic orders \citep{Xu} and the transition between BCS (Bardeen--Cooper--Schrieffer model) and BEC states (Bose--Einstein condensation) \citep{Gazit} have also been observed in these models. However, none of above works show the effective interaction of Ising spins at the weak coupling limit between fermion and Ising spin.

	 The Ruderman--Kittel--Kasuya--Yosida (RKKY) interactions between localized spins are formed indirectly via nonlocal electrons \citep{Ruderman, Kasuya, Yosida}. That is one of the most important prototypes to explain the formation of magnetic order in pure rare-earth elements (e.g Gd, Sm, and Dy), their alloys, heavy fermion materials, diluted magnetic semiconductors, and impurities in graphene \citep{Kondo1, Kondo2, Uchoa, Lee, Kotov}. Therefore, the derivation of RKKY interaction is varied from one system to the other with a main spin susceptibility, or static Lindhard function in the momentum space $\vec{q}$ and  the real space $\vec{R}$:
\begin{align}
\centering
\chi(\vec{q}) & = \int_{\vec{k} \in \text{BZ}} \frac{n_F(E_{\vec{k}}) - n_F(E_{\vec{k} +\vec{q}})}{ E(\vec{k} + \vec{q}) - E(\vec{k})}, \label{eqn11}\\
\chi(\vec{R}) & = \int_{\vec{q} \in \text{BZ}} \frac{(d\vec{q})^d}{(2 \pi)^d} e^{-i \vec{q} \cdot \vec{R}} \chi(\vec{q}). \label{eqn22}
\end{align}  
	In the Eq.~(\ref{eqn11}), the momentum space of Lindhard function $\chi(\vec{q})$ is calculated by integrating over the first Brillouin zone (BZ), where $n_F(E_{\vec{k}})$ is the Fermi-Dirac distribution function, and $E(\vec{k})$ is the energy dispersion relation in the momentum $\vec{k}$. The spin susceptibility consists the singular or maximum point in the momentum space that define the magnetic ordering vector of spin system. The real-space susceptibility $\chi(\vec{R})$ is achieved by taking the Fourier transformation of the momentum space one with spatial dimensions $d$ Eq.~(\ref{eqn22}). The real-space $\chi(\vec{R})$ function is oscillatory and decaying with the distance $R$ of two localized spins. For example, the conventional real RKKY interaction of spins showed the sign-changing oscillation and decaying rate of $1/R^d$ ($d$ is the dimension of crystal lattice) \citep{Fischer}. However, real RKKY interaction between magnetic impurities in the graphene shows no sign-changing oscillation due to vanishing of the Fermi surface, and the $1/ R^3$ decaying rate instead of $1/ R^2$ for 2D \citep{Kotov, Uchoa}. The microscopic magnetic interactions of impurities in the graphene are known as FM and AFM couplings for spins on the same and different sublattices, respectively. The exact macroscopic magnetic order of impurity-doped graphene is quite controversial discussion such as AFM \citep{Annica, Brey} and FM orders \citep{Saremi}. 
	
	This letter, we have constructed the total Hamiltonian, including fermions at the sites and Ising spins at the bonds, of the 0-flux and $\pi$-flux square lattices. We solve problem by two different approaches: semi-analytic integration and exact diagonalization. In the semi-analytic method, we separate the Hamiltonian to unperturbing (pure hopping between fermions) and interacting parts (coupling between fermion and Ising spin). Using the perturbation theory, the interacting Hamiltonian is treated at the weak limit. The first- and second-order coupling terms are derived analytically, then computed numerically in the momentum and real spaces. Since the frustrating effect emerges in our models due to competing effect between nearest and next-nearest couplings, the exact diagonalization method is used to verify semi-analytic calculations. Based on microscopic discussions, we will determine the final magnetic order and the ground-state energy for each case.
\section{Model and Methods}
\subsection{Model}
\begin{figure}[!ht]  
  \centering
  \subfloat[]{\includegraphics[width=0.47\textwidth]{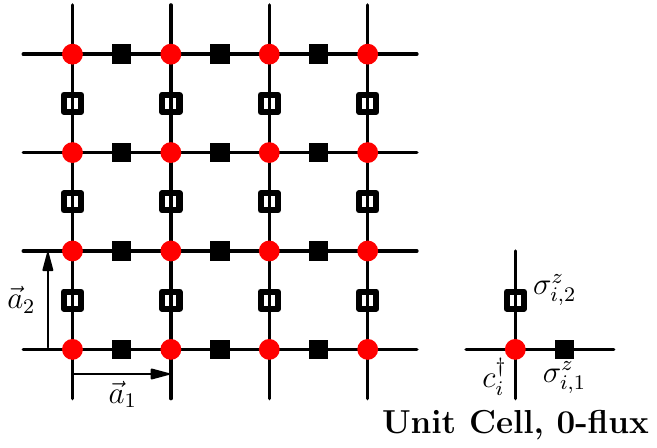}\label{fig:f1a}}
  \hfill
  \subfloat[]{\includegraphics[width=0.48\textwidth]{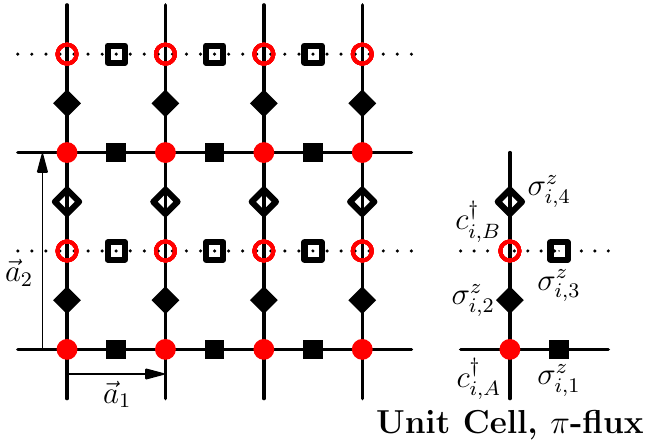}\label{fig:f1b}}
  \caption{\small (a) Each unit cell $i$ of the 0-flux square lattice consists one fermion $c^{\dagger}_{i}$ at the site, and two Ising spins $\sigma^{z}_{i,1}$ and $\sigma^{z}_{i,2}$ arranged along the x- and y-directions, respectively. The solid line illustrates the nearest-neighbor hopping integral $-t$ between fermions. Two lattice vectors are $\vec{a}_1 = (1, 0)$ and $\vec{a}_2 = (0, 1)$, here we consider the lattice constant unit. (b) The unit cell of $\pi$-flux lattice is doubled in the size of 0-flux one. It includes two distinct fermions $c^{\dagger}_{i, A}$ and $c^{\dagger}_{i, B} $ and four Ising spins such as $\sigma^z_{i, 1}$, $\sigma^z_{i, 2}$, $\sigma^z_{i, 3}$ and $\sigma^z_{i, 4}$. Solid lines represent the hopping integral $-t$, and dotted lines mean $+t$ hopping integral between the adjacent orbital fermions. Two lattice vectors are $\vec{a}_1 = (1, 0)$ and $\vec{a}_2 = (0, 2)$.  } \label{fig:f1}
\end{figure}
%
The general Hamiltonians $\hat{H}$ of the 0- and $\pi$-flux square lattices are formed:
%
\begin{equation}
\begin{split}
 \hat{H} &= \sum_{\langle m, n \rangle } -t_{ m n}( 1 + \xi \sigma_{ m n }^{z}) (c^{\dagger}_{ m } c_{ n} + c^{\dagger}_{ n } c_{  m })\\
 & = -\sum_{\langle m,n \rangle} t_{mn} (c^{\dagger}_{m} c_{n} + c^{\dagger}_{n} c_{m})\\
 &\hspace{2cm}  -\sum_{\langle m,n \rangle}  t_{mn} \xi \sigma_{mn}^{z} (c^{\dagger}_{m} c_{n} + c^{\dagger}_{n} c_{m})\\
 &\equiv \hat{H}_0 + \hat{H}_1.
\end{split} 
\end{equation}

	Here, spin-1/2 fermion $c^{\dagger}_{m} = (c^{\dagger}_{m \uparrow}, c^{\dagger}_{m \downarrow}) $ lives at the site $m$ of the square lattice, and $\langle m,n \rangle$ is the nearest-neighbor pair of the orbital fermions. The Ising spin $\sigma_{mn}^{z}  $ is positioned at the bond between two fermions $m$ and $n$ -- link-centered $\text{Z}_2$ model.  For the 0-flux lattice, there are two Ising spins in each unit cell $i$: $\sigma^z_{i, 1} $ and $\sigma^z_{i, 2} $ (Fig.~\ref{fig:f1a}). We set the hopping integral $t_{mn} = t$ and the coupling parameter $ t_{mn} \xi$ ($ t$ and $\xi$ are real positive values). For the $\pi$-flux model, there are four Ising spins in each unit cell: $\sigma^z_{i, 1} $, $\sigma^z_{i, 2} $, $\sigma^z_{i, 3} $ and $\sigma^z_{i, 4} $  (Fig.~\ref{fig:f1b}). Along the solid line in the Fig.~\ref{fig:f1b}, $t_{mn} = t$, and along the dotted line $t_{mn}$ is reversed sign due to the $\pi$-flux passing through each plaquette. (Note: in 0-flux lattice, the number of site equals to number of unit cell, so $i = m$. However, in $\pi$-flux lattice, the number of site is doubled number of unit cell, so two values are different.)
	
	We separate the full Hamiltonian into the tight-binding part $ \hat{H}_{0} $ and interacting part $ \hat{H}_{1} $ with the small perturbation variable $\xi$. The Hamiltonian is written explicitly in the real space with the lattice translation vector $\vec{R}$, then transformed into the momentum space $\vec{k}$ (see Appendix~A). In our model, the interacting term represents coupling between Ising spins and the itinerant fermions compared with the spin of fermions and localized spins in Kondo lattice -- site-centered SU(2) case: 
\begin{align}
 \hat{H}_{\text{K}} & = -\sum_{\langle m,n \rangle} t_{mn} (c^{\dagger}_{m} c_{n} + c^{\dagger}_{n} c_{m})  - \xi \sum_{m}  \vec{s}_m \cdot \vec{S}_m,  \label{sitemodel} 
\end{align}
where $\vec{s}_m$ is spin of itinerant electron, and $\vec{S}_m$ is the localized spin at the position $m$.
This term is used to derive the second-order conventional spin susceptibility of RKKY interaction \citep{Budapest}.

	Below, we show the main steps  of two calculating methods.
\subsection{Semi-analytic integration}	
\subsubsection{Momentum-dependent 0-flux model} 
	The momentum-dependence of the non-interacting Hamiltonian $\hat{H}_0$ is diagonalized to give the energy dispersion $E^{0}_{0-\text{flux}}(\vec{k})$:
\begin{align}
E^{0}_{0-\text{flux}}(\vec{k}) & = -2t (\cos{k_x} + \cos{k_y}).   \label{Fermi dispersion} 
	\end{align}
	In the 0-flux model, the expectation energies of the interacting Hamiltonian $\hat{H}_{1}(\vec{k},\vec{q})$ are calculated using perturbation theory with the Fermi-sea ground state. The first-order energy is found:
\begin{multline}\label{1-zero}
E^{1}_{0-\text{flux}} (\vec{q})= \frac{-\xi}{N} \sum_{\vec{k}}  2 n_F(E_{\vec{k}})\\
\times\Big [ \cos{k_x}  
\sigma^{z}_{\vec{q}, 1} +  \cos{k_y} \sigma^{z}_{\vec{q}, 2} \Big ] \delta_{\vec{q}, \vec{0}}, 
\end{multline} 
where $N$ is number of unit cells in the lattice, and $n_F(E_{\vec{k}}) $ is Fermi--Dirac function.
	The second-order perturbation provides a $2\times 2$ interacting matrix $ J^\text{RKKY}_{\alpha \beta}(\vec{q}, \omega_n) $ inside the effective energy $E^{2}_{0-\text{flux}} (\vec{q})$:
%
\begin{align} 
\centering
E^{2}_{0-\text{flux}} (\vec{q}) & = \sum_{\alpha, \beta = 1}^{2} \sigma^{z}_{\vec{q}, \alpha} J^\text{RKKY}_{\alpha \beta}(\vec{q}, \Omega_n) \sigma_{-\vec{q},\beta}^z, \label{H_Fermi}
\\           
 J^\text{RKKY}_{\alpha \beta}(\vec{q}, \Omega_n) & = -\frac{\xi^2}{N^2} \sum_{\vec{k} \in \text{BZ}} J_{\alpha \beta}(\vec{k}, \vec{q}) \chi^{\text{F}}(\vec{k}, \vec{q}, \omega_n), \\ 
\chi^\text{F}(\vec{k}, \vec{q}, \Omega_n) & =  \frac{n_F(E_{\vec{k}}) - n_F(E_{\vec{k} + \vec{q}})} {i\omega_n + E_{\vec{k} + \vec{q}} - E_{\vec{k}}}.                                       
\end{align} 
Here, $\alpha$ and $\beta$ are labeled of Ising spin in the unit cell of 0-flux lattice (Fig.~\ref{fig:f1a}). $\chi^{\text{F}}(\vec{k}, \vec{q}, \omega_n)$ is the Lindhard function for Fermi metallic band or intra-band interaction. All elements of $J_{\alpha \beta}(\vec{k}, \vec{q}) $ matrix are listed in the Sect.~1 of Appendix~B. 

	To compute $\vec{q}$-space RKKY interaction, we take integration over the whole square Brillouin zone for momentum $\vec{k}$ and diagonalize $J^\text{RKKY}_{\alpha \beta}(\vec{q}, \omega_n)$ matrix to obtain two eigenvalues \textit{Eigen1} and \textit{Eigen2}. We divide $k_x$ and $k_y$ from $-\pi$ to $\pi$ into $N$ intervals, and $N = L_x = L_y = 400$ ($L_x$ and $L_y$ are the lengths of real lattice). For Matsubara frequency $\Omega_n = 2nT$, we select the value of $\Omega_n / T = 10^{-5}$ and integer number $n$. Two eigevalues as the functions of $\vec{q}$ are plotted along high symmetric points in the Brillouin zone such as $\Gamma = (0,0)$, $M = (\pi, 0)$ and $K = (\pi, \pi)$ in Fig.~\ref{fig:f2a}. 
\subsubsection{Momentum-dependence of the $\pi$-flux model}

	We transform the unperturbed Hamiltonian $\hat{H}_0$ into the momentum space and diagonalize it to obtain the energy dispersion with two bands:
\begin{equation} \label{Dirac-band}
 E_{(2,1), \pi-\text{flux}}^{0} (\vec{k})  = \pm \big [ 2t (\cos^2{k_x} + \cos^2{k_y} ) \big ] ^{1/2}.
\end{equation}
%

	Similar to the 0-flux model, the interacting Hamiltonian of $\pi$-flux model gives the first- and second-order effective energies. The first-order is:
\begin{equation} 
\begin{split}
\label{1-pi}
E^{1}_{\pi-\text{flux}} (\vec{q}) & = \frac{-\xi}{N} \sum_{\vec{k}}   n_F(E_{1, \vec{k}})\Bigg [  \frac{2\cos{k_x} u^{2}(\vec{k})}{\varv^2(\vec{k})}   \sigma^{z}_{\vec{q}, 1}
\\
&\quad \frac{[ 1+ \cos{2k_y} ] u(\vec{k})}{\varv^2(\vec{k})}  ( \sigma^{z}_{\vec{q}, 2} + \sigma^{z}_{\vec{q}, 4} )
\\
&\quad \frac{\cos{k_x} [ 1+ \cos{2k_y} ]}{\varv^2(\vec{k})}   \sigma^{z}_{\vec{q}, 3} \Bigg ] \delta_{\vec{q}, \vec{0}}. 
\end{split}
\end{equation}
Here, $u(\vec{k})$ and $\varv(\vec{k})$ functions are defined in the Sect.~2 of Appendix~B.
	 
	The second-order effective energy $E^{2}_{\pi-\text{flux}} (\vec{q})$ of the interacting Hamiltonian includes a $4 \times 4$ matrix with common Lindhard function:
\begin{align} 
E^{2}_{\pi-\text{flux}} (\vec{q}) & = \sum_{\alpha, \beta = 1}^{4} \sigma^{z}_{\vec{q}, \alpha} J^\text{RKKY}_{\alpha \beta}(\vec{q}, \omega_n) \sigma_{-\vec{q},\beta}^z,            
\\
J^\text{RKKY}_{\alpha \beta}(\vec{q}, \omega_n) & =  -\frac{\xi^{2}}{N^2} \sum_{\vec{k} \in \text{BZ}} M_{\alpha \beta}(\vec{k}, \vec{q})
              \chi^{\text{D}}(\vec{k}, \vec{q}, \omega_n), \label{RKKY-piflux}
\\               
\chi^{\text{D}}(\vec{k}, \vec{q}, \omega_n) & =  \sum_{s, s^{\prime}}             
\frac{n_F(E_{\vec{k}, s}) - n_F(E_{\vec{k} + \vec{q}, s^{\prime}})} {i\omega_n + E_{\vec{k} + \vec{q}, s^{\prime}} - E_{\vec{k}, s}}.               
\end{align}
	Here, $\alpha$ and $\beta$ are labeled of Ising spins in the unit cell of $\pi$-flux lattice (Fig.~\ref{fig:f1b}). $\chi^{\text{D}}(\vec{k}, \vec{q}, \omega_n) $ is the Lindhard function for Dirac semimetallic bands. The values of $s$ and $s^{\prime}$ represent the energy bands. For the half-filling case, there is only one situation with $s = 1$ and $s^{\prime} = 2$ results in the non-zero value of Lindhard function, or it is interband interaction \citep{Brey}. (All the terms of matrix $M_{\alpha \beta} (\vec{k}, \vec{q})$ are in the Sect.~3 of Appendix~B.) 
	
	We find the eigenvalue spectrum of $J^\text{RKKY}_{\alpha \beta}(\vec{q}, \omega_n) $ by taking integration over the rectangular Brillouin zone and diagonalizing $4 \times 4$ functional matrix. Analogous to the 0-flux model, that function shows four different eigenvalues such as \textit{Eigen1}, \textit{Eigen2}, \textit{Eigen3} and \textit{Eigen4} that plot along the high symmetric points such as $\Gamma = (0, 0)$, $X = (\pi, 0)$, $Y = (0, \pi/2)$, and $K = (\pi, \pi/2)$ in Fig.~\ref{fig:f2b}.	
\subsubsection{Real-space calculation}	
	Previous formulas are written in the momentum space. We compute numerically the real-space RKKY interaction between Ising spin pair by taking the Fourier transformation of the momentum space of $J_{\alpha \beta}(\vec{q}, \omega_n) $ function:
\begin{equation} \label{equation 3.4}
J_{\alpha \beta}^{\text{RKKY}}(\vec{R})  = \int_{\vec{q} \in \text{BZ}} \frac{dq_x dq_y }{(2 \pi)^2} e^{-i\vec{q} \cdot \vec{R}} J_{\alpha \beta}(\vec{q}, \omega_n),    
\end{equation}
where $\vec{R} = \vec{R}_j - \vec{R}_i$ is the distance vector of two $i$ and $j$ unit cells. For the 0-flux lattice, the $\alpha, \beta = 1$ and 2 (or Ising spins 1 and 2). For the $\pi$-flux lattice, the $\alpha, \beta = 1, 2, 3$ and 4 (or Ising spins 1, 2, 3 and 4). All interacting functions are plotted via the real distance $R = | \vec{R}|$ along the x- or y-directions.  We use the lattice size of $L_x = L_y = 160$, and $L_x = 160$ and $L_y = 80$ for 0- and $\pi$-flux models, respectively. From those calculations, we have constructed the real-space effective interacting energy of the  system.
\subsection{Exact diagonalization}
	 Following the illustrated square lattices in the  Fig.~\ref{fig:f1}, the real Hamiltonian matrices are constructed based on the hopping amplitude $t$ and coupling parameter $\xi$ with their sizes of $N = L_x \times L_y = 160 \times 160$. With the 0-flux model, the hopping terms along the x- and y-directions are $-t - \xi \sigma^z_{1}$ and $-t - \xi \sigma^z_{2}$, respectively. We set the value of $t = 1$, and the total energy $E^{\text{total}} = \langle \Psi_{\text{gs}} | \hat{H} | \Psi_{\text{gs}} \rangle$ depends on the Ising configuration and parameter $\xi$. 
	
	Since the lattice size is large $L_x \times L_y = 160 \times 160$, the number of Ising spin in this lattice is $ N_{\text{Ising}} = 51200$ spins. It is impossible to find the minimal energy of the system by optimizing over all Ising spin configurations. Because the definition of the magnetic ordering vector $\vec{Q} $ would be at the high symmetric points of Brillouin zone, we have found the period of Ising spin configuration using formula $\vec{Q} \cdot \vec{r}_{m} = 2\pi m $ with integer number $ m$ \citep{Legg, She}. For example, in the 0-flux model, if we choose the magnetic ordering vector is at the M point in Fig.~\ref{fig:f2a}, $\vec{Q} = (\pi, 0)$, and set up Ising 1: spin-up (or $\sigma^z_{1} = 1$) and Ising 2: spin-down (or $\sigma^z_{1} = -1$), all the spins are reversed directions at next unit cell on the right. Along the y-direction, they copy the similar configuration. The obtaining result of numerical method is used to compare with the semi-analytic calculation one to determine the correct spin orders for both models.  
\section{Results and Discussions}
\subsection{Energy dispersions}
	\begin{figure}[!ht]
\centering
  \subfloat[]{%
    \includegraphics[width=.263\textwidth]{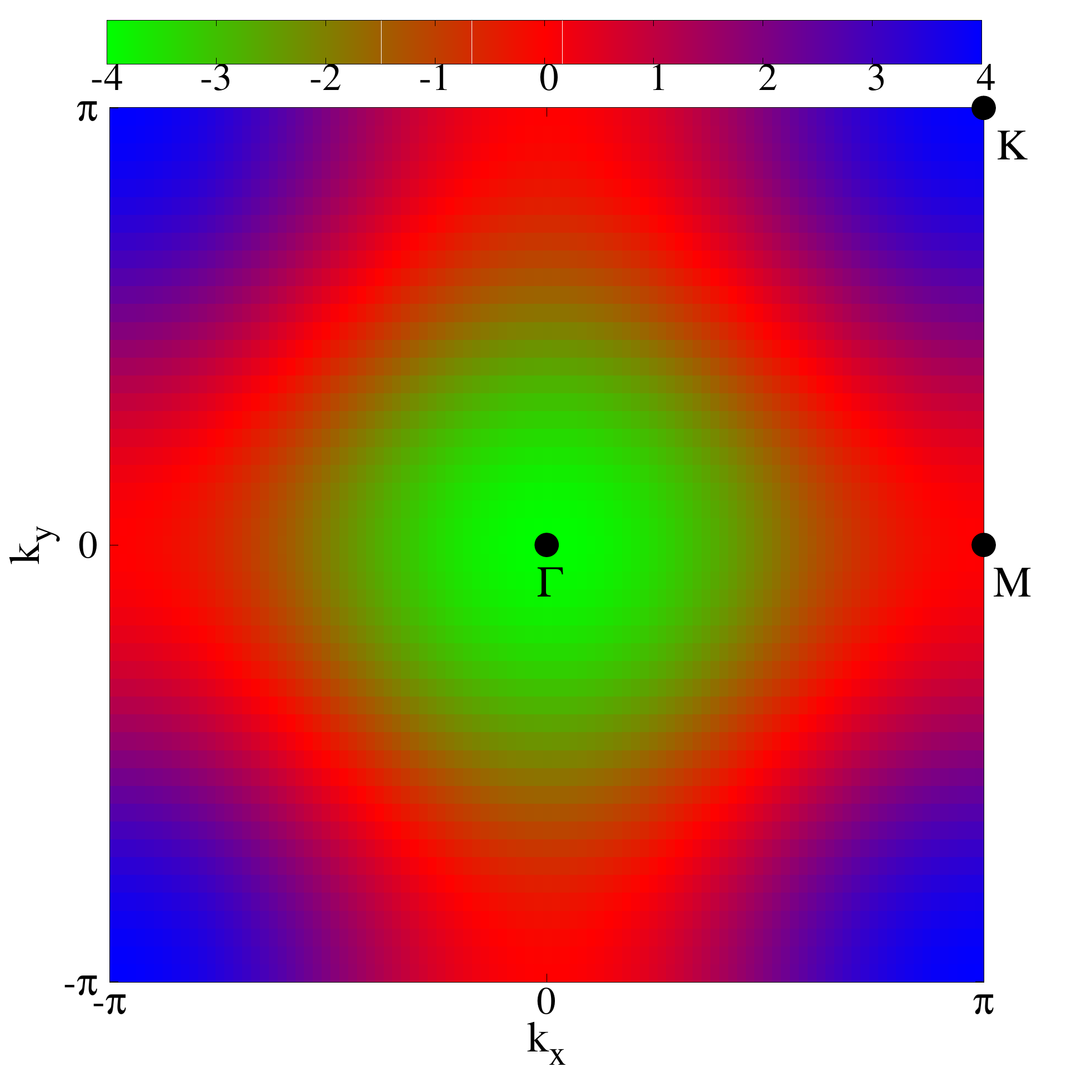}\label{fig:f2a}     \hfill }
   \subfloat[]{%
    \hspace{-0.3cm}\includegraphics[width=.263\textwidth]{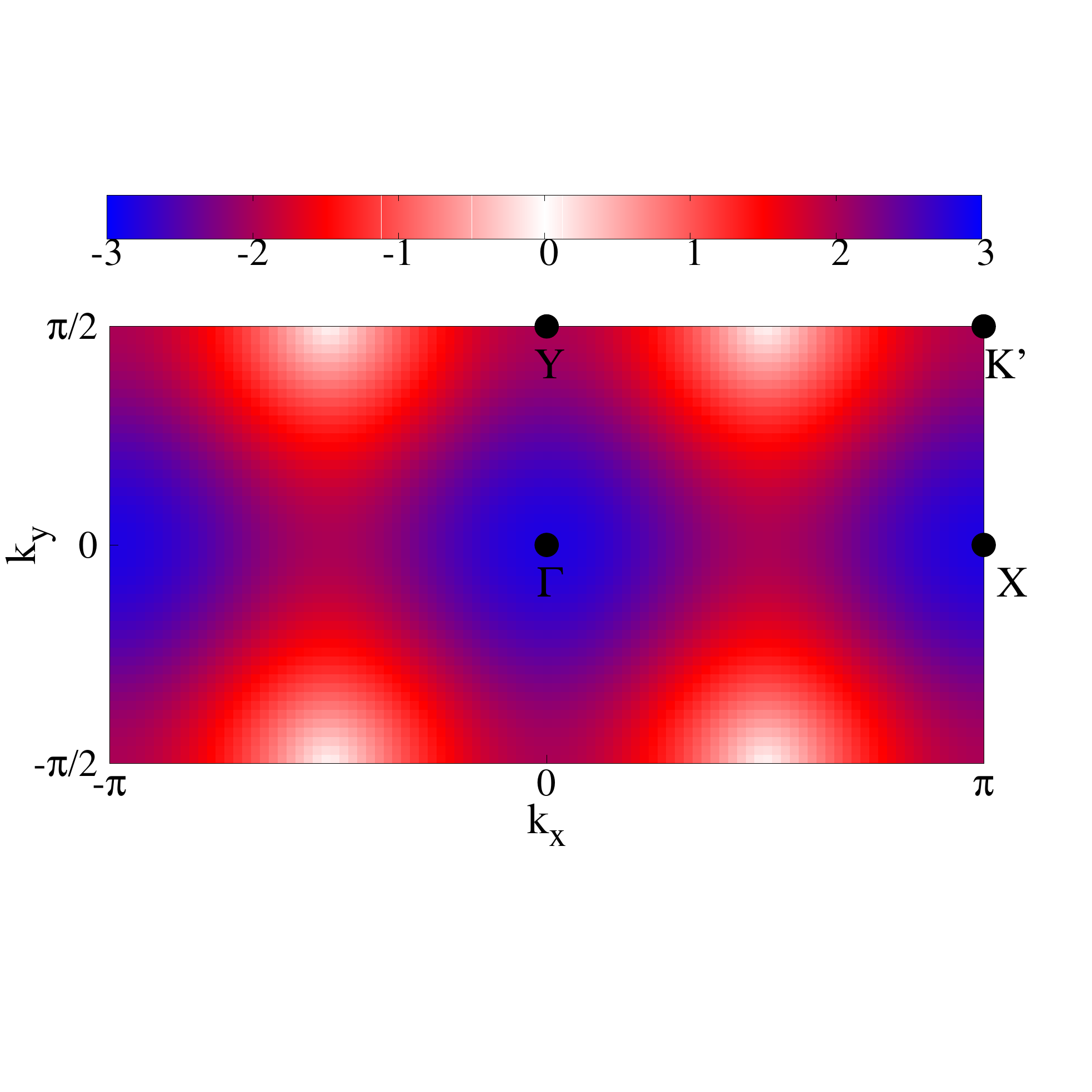}\label{fig:f2b}}
 \caption{\small Two-dimensional contour plots of (a) the 0-flux lattice with the energy dispersion $E^{0}_{0-\text{flux}} = -2 (\cos{ k_x} + \cos{ k_y} )$, and (b) the $\pi$-flux lattice with the energy dispersions $E_{(2,1), \pi-\text{flux}}^{0} = \pm \big [ 2 (\cos ^2{k_x} + \cos^2 {k_y} ) \big ]^{1/2} $ (the hopping amplitude $t = 1$ refers to equations (\ref{Fermi dispersion}) and (\ref{Dirac-band})).} \label{fig2}
\end{figure}
	Figure \ref{fig2} shows the 2D contour plots of Fermi band and Dirac bands for the 0-flux and $\pi$-flux lattices, respectively. The energy dispersion $E^{0}_{0-\text{flux}} = -2 (\cos{ k_x} + \cos{ k_y} )$ of 0-flux lattice exhibits the continuous metallic band and diamond shape Fermi surface (red line in Fig.~\ref{fig:f2a}). (The 3D surface is also plotted in my MSc dissertation \citep{HDo}). At the half-filling or chemical potential $\mu = 0$, each site of lattice is occupied exactly one electron. The Fermi surface of the 0-flux square lattice provides nesting property. There are nesting vector $\vec{Q}_0 = (\pm \pi, \pm \pi)$ that connect all points on the Fermi surface \citep{Budapest}. The high symmetric points in this Brillouin zone include $\Gamma = (0,0)$, $\text{M} = (\pi, 0) $, and $\text{K} = (\pi, \pi) $.

	2D contour of the $\pi$-flux lattice is plotted with the energy dispersions $E_{(2,1), \pi-\text{flux}}^{0} = \pm \big [ 2 (\cos ^2{k_x} + \cos^2 {k_y} ) \big ]^{1/2} $. The upper $E_2^0$ and lower $E_1^0$ bands contact each other at four Dirac points $D = (\pm \pi / 2, \pm \pi / 2$) (white region in the  Fig.~\ref{fig:f2b}). So, the upper band  $E_{2}^{0}$ is empty state (electron band), and lower band  $E_{1}^0$ is completely filled (hole band). That is a typical band structure of semimetal (an example of graphene \citep{C_Neto}). Four distinct symmetric points in the rectangular Brillouin zone are $\Gamma = (0,0)$, $\text{X} = (\pi, 0) $, $\text{Y} = (0, \pi/2) $, and $\text{K}^{\prime} = (\pi, \pi/2) $. 	
\begin{figure}[!ht]
  \centering
  \subfloat[]{\includegraphics[width=6.5cm, height=6.5cm]{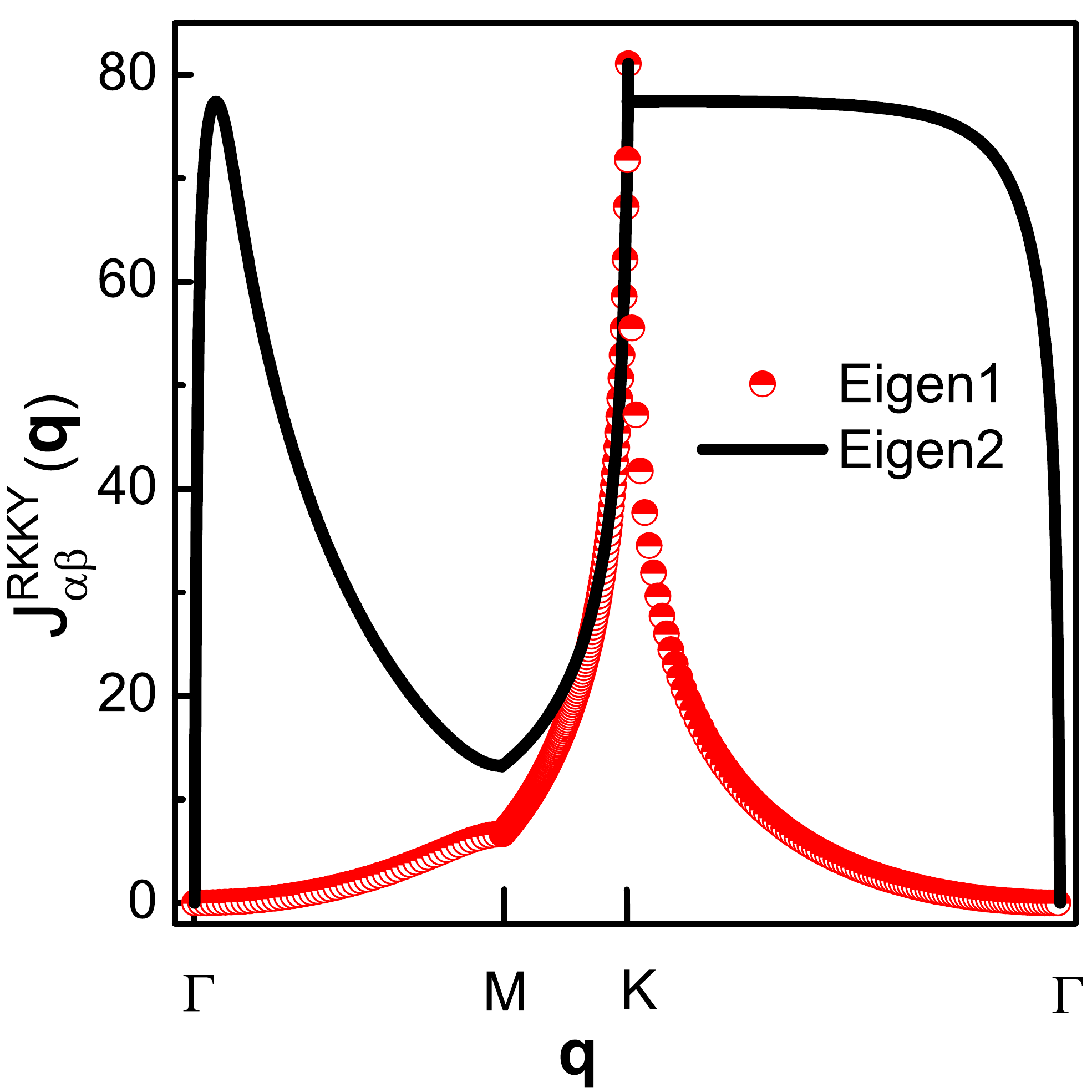}\label{fig:f3a}}
  \hfill  
  \subfloat[]{\includegraphics[width=6.5cm, height=6.5cm]{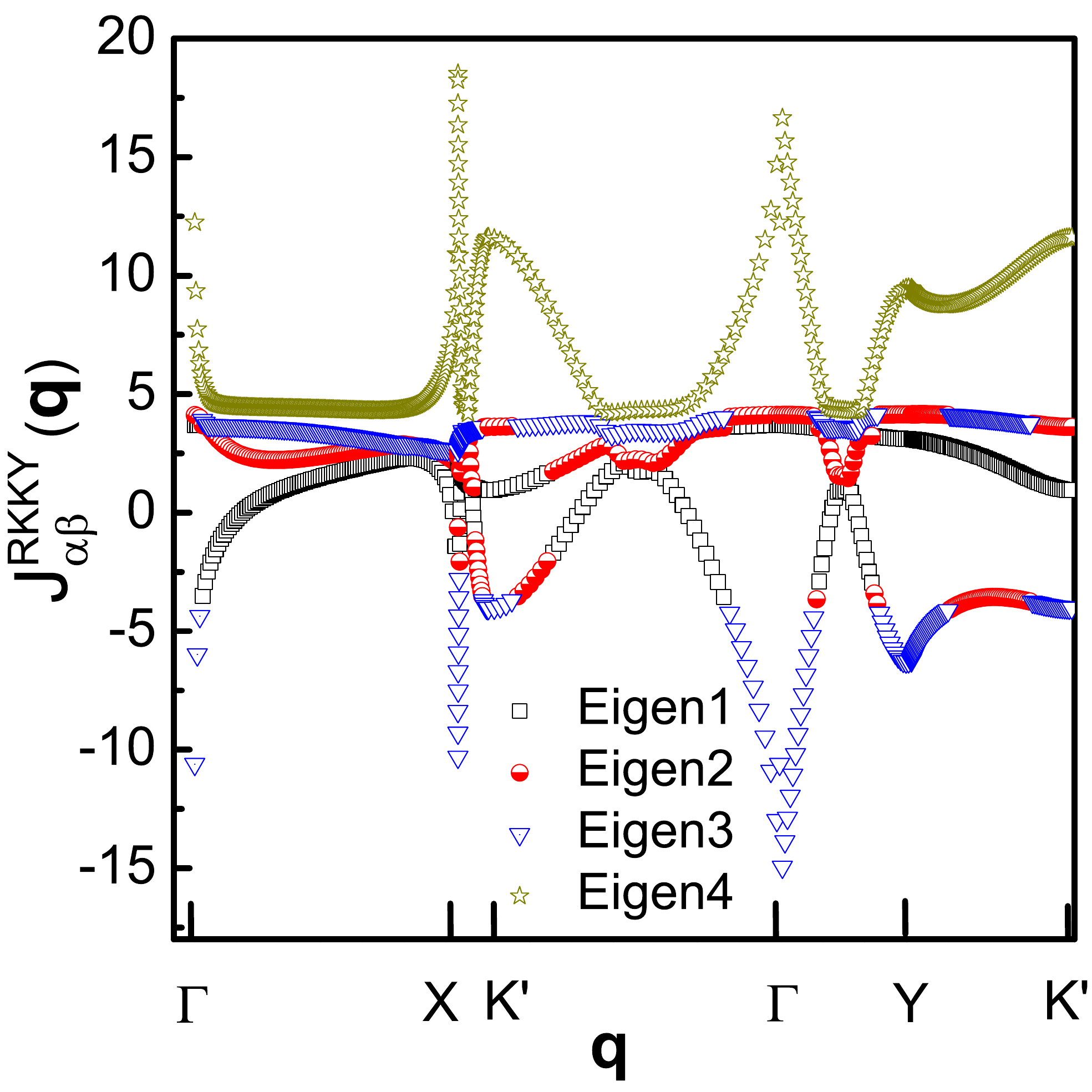}\label{fig:f3b}}
  \caption{ \small The momentum-dependence of  $J^{\text{RKKY}}_{\alpha \beta}(\vec{q})$ function (a) in 0-flux model plotting along $\Gamma$--M--K--$\Gamma$ path in the square Brillouin zone ($\Gamma = (0,0)$, $\text{M} = (\pi, 0)$ and $\text{K} = (\pi, \pi)$ in Fig.~\ref{fig:f2a}) and (b) in the $\pi$-flux model plotting along $\Gamma$--X--$\text{K}^{\prime}$--$\Gamma$--Y--$\text{K}^{\prime}$ path in the rectangular Brillouin zone ($\Gamma = (0,0)$, $\text{X} = (\pi, 0)$, $\text{K}^{\prime} = (\pi, \pi/2)$ and $\text{Y} = (0, \pi/2) $ in Fig.~\ref{fig:f2b}). }
  \label{fig3}
\end{figure}
\begin{figure*}[!ht]
  \centering
  \subfloat[]{\includegraphics[width=0.30\textwidth]{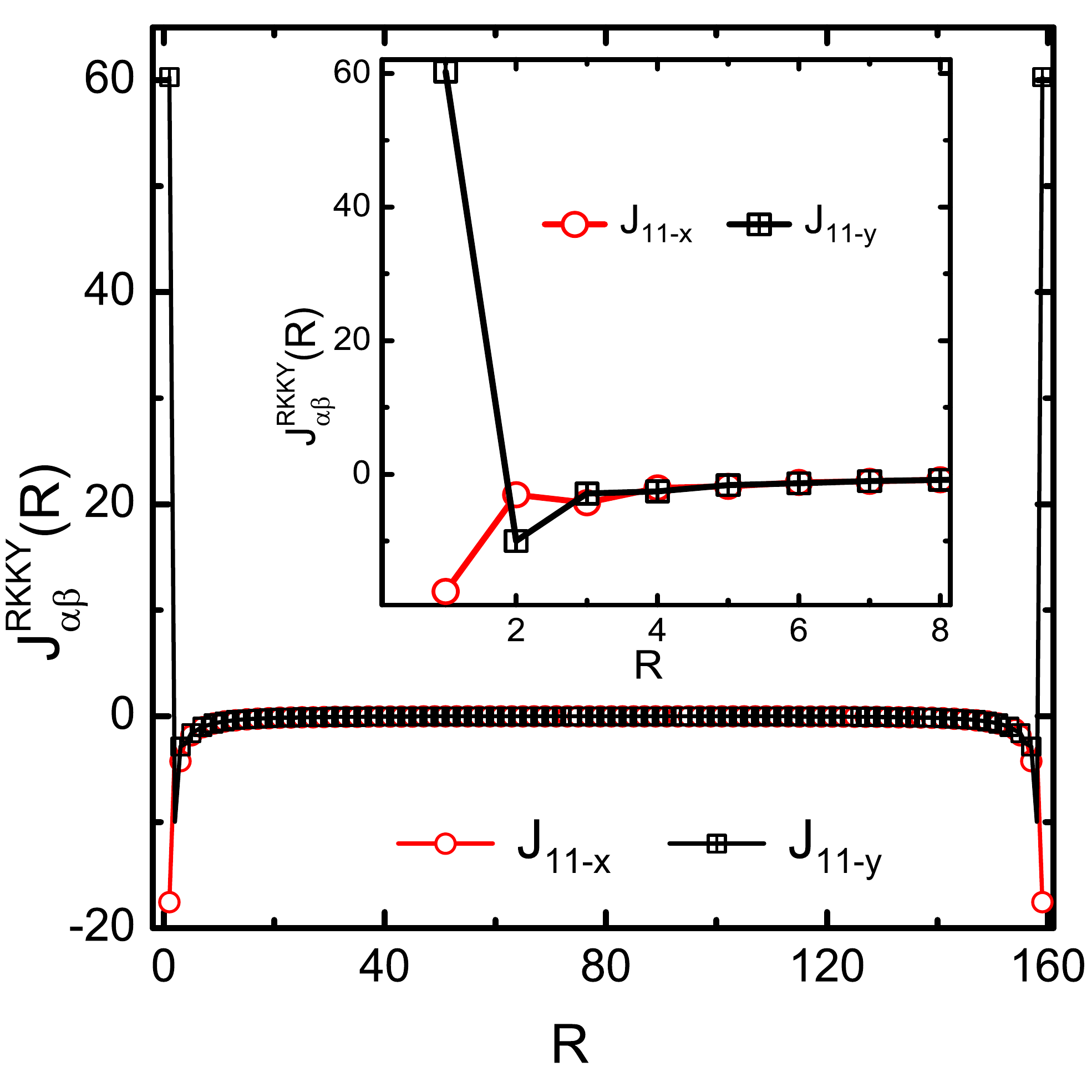}\label{fig:f4a}  \hfill}
  \subfloat[]{\includegraphics[width=0.30\textwidth]{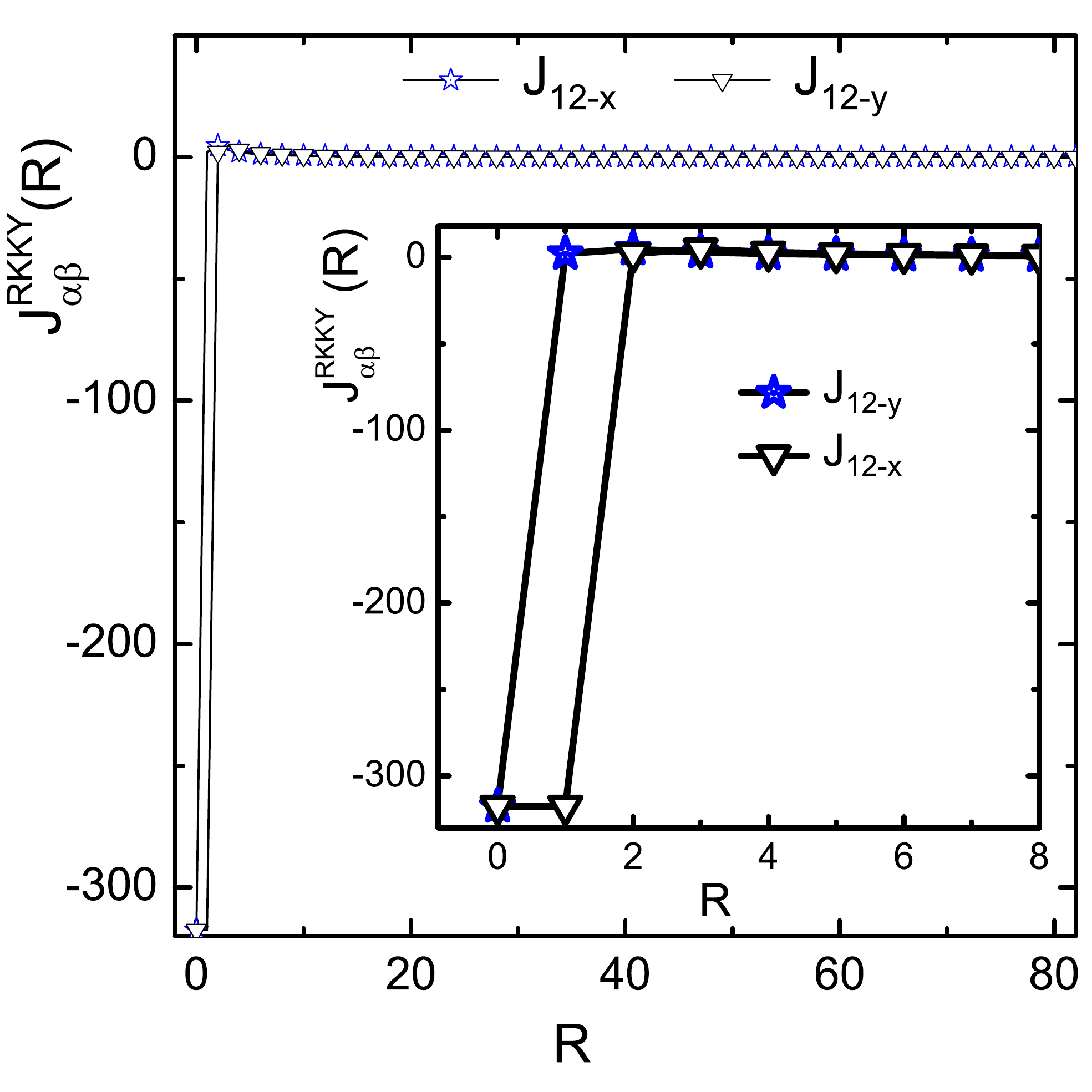}\label{fig:f4b} \hfill}
  \subfloat[]{\includegraphics[width=0.30\textwidth]{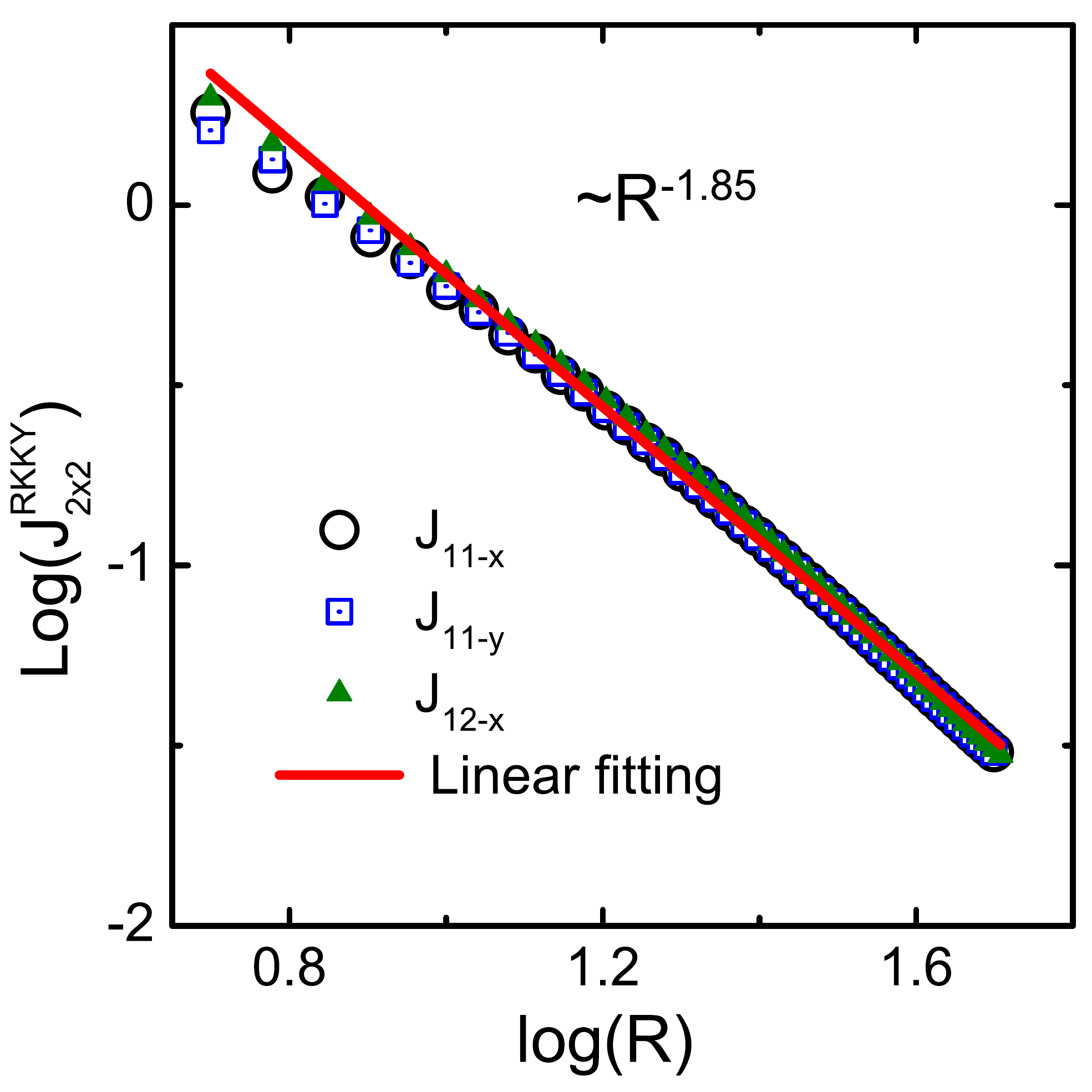}\label{fig:f4c}} 
  \caption{\small Real-space dependence of (a) $J_{11}^{\text{RKKY}}(R)$ (the interaction between two pairwise Ising spins 1 and 1) (b) $J_{12}^{\text{RKKY}}(R)$ (the interaction between pairwise spins 1 and 2) along the x- and y-directions with the lattice size $L = 160$ (where $R = \abs{\vec{R}} = \abs{\vec{R}_j - \vec{R}_i}$ is the distance of spins in two unit cell $i$ and $j$ for the x- and y-directions). The insets of figures zooms in the effective interactions between Ising spins within 8 unit cells. (c) $\log J_{\alpha \beta}^{\text{RKKY}}$ -- $\log R $ relation of those interacting spin pairs.}
  \label{fig4}
\end{figure*}	
\subsection{Eigenvalue spectra in the momentum space} 
%
	Figure~\ref{fig3} shows the momentum-dependence of the second-order susceptibility of the 0-flux and $\pi$-flux models. The eigenvalue spectrum of $J^{\text{RKKY}}_{\alpha \beta}(\vec{q})$ ($\alpha, \beta = $ 1 and 2) matrix is plotted along the high symmetric points of the square zone such as $\Gamma = (0,0)$, $\text{M} = (\pi, 0)$ and $\text{K} = (\pi, \pi)$. Because of symmetric properties between the x- and y- directions, we only calculate it at the M$(\pi, 0)$ point. That spectrum shows degenerate maximum value at the K = ($\pi$, $\pi$) point. According to the She's discussions \citep{She}, the ordering vector of magnetic interaction should be defined at the maximum point K of eigenvalue plot. That is due to the maximum value with minus sign of coupling constant $-\xi^2/N^2$ gives the minimum magnetic energy or stable system (see equations \ref{H_Fermi} and \ref{RKKY-piflux}) \citep{She}. However, there is the other hidden singular point which does not show in the way we calculate eigenvalue spectrum. RKKY interaction is called the static function as $\omega_n = 0 $ and $\vec{q} \to \vec{0}$. The $J_{11}(\vec{q})$, for example, is:
\begin{eqnarray}
J_{11}(\vec{q} \to \vec{0}, \omega_n = 0)	 & = & 	\frac{2}{\pi}	\int_{-\pi}^{0} dk_x \frac{\cos^2(k_x)}{\sin(k_x)}.		
\end{eqnarray}
The final integral form of $J_{11}(\vec{q} \to \vec{0}) $ is diverged at those integral limits. It is an approximation we compute the RKKY interaction in the 0-flux square lattice (see the Sect.~1 of Appendix~C for detail calculation).   
	
	The eigenvalue spectrum of $J^{\text{RKKY}}_{\alpha \beta}(\vec{q})$ ($\alpha, \beta = $ 1, 2, 3 and 4) matrix of $\pi$-flux model is showed in  Fig.~\ref{fig:f3b}. Four distinct eigenvalue curves includes $Eigen1$, $Eigen2$, $Eigen3$ and $Eigen4$ are plotted along $\Gamma = (0,0)$, $\text{X} = (\pi, 0)$, $\text{K}^{\prime} = (\pi, \pi/2)$ and $\text{Y} = (0, \pi/2)$. The $Eigen4$ plot shows the highest value at the $\text{X} (\pi, 0)$ point. Since the Lindhard function of $\pi$-flux model is interband transition, there is no appearance of zero value at its numerator and denominator. The maximum point defines the magnetic order $\vec{Q} = (\pi, 0)$ of the system.
\subsection{Real-space semi-analytic calculations}
\begin{figure}[htb]
  \centering
 \hspace*{\fill}  
  \subfloat[]{\includegraphics[width=0.248\textwidth]{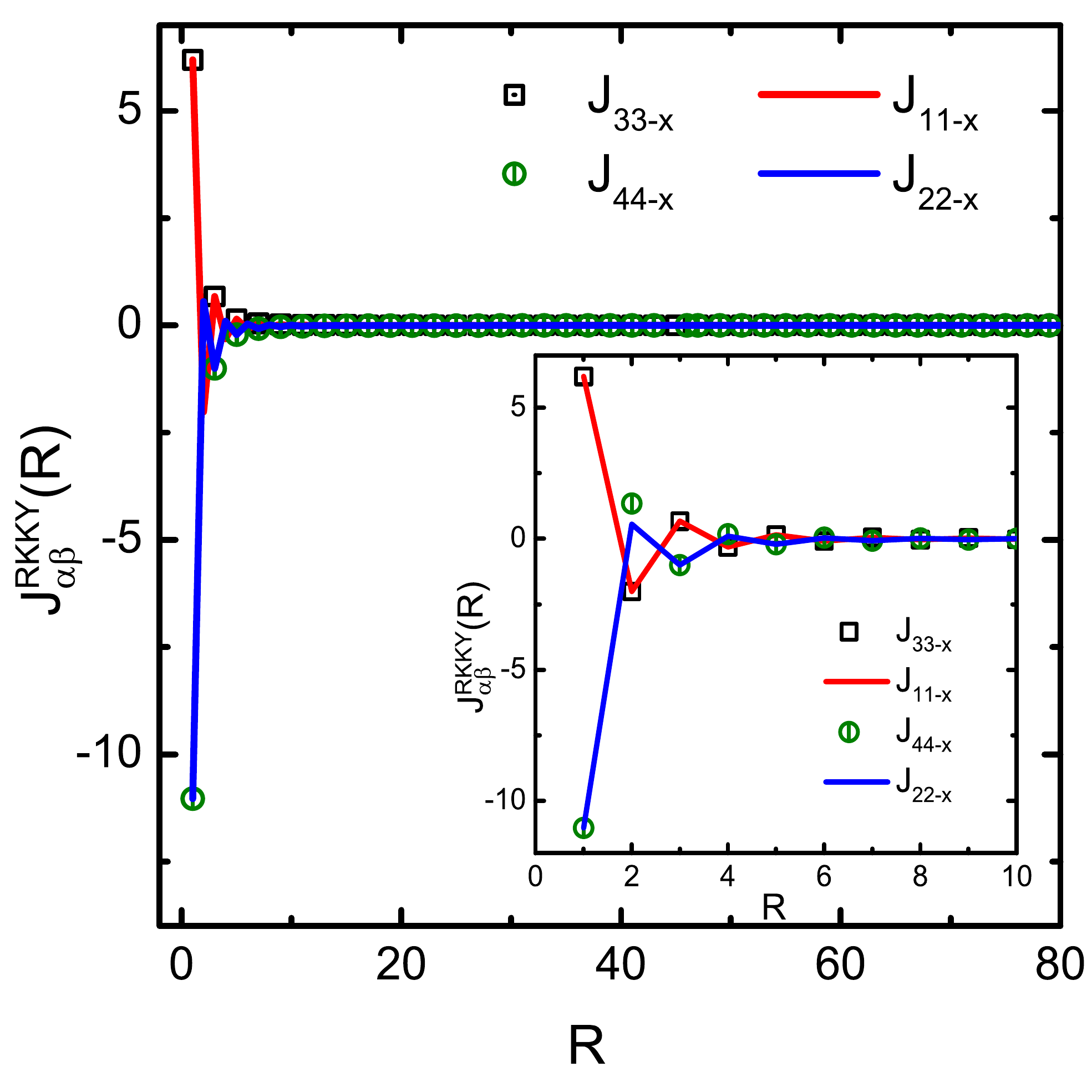}\label{fig:f5a} \hfill}
  \subfloat[]{\hspace{-0.13cm}\includegraphics[width=0.248\textwidth]{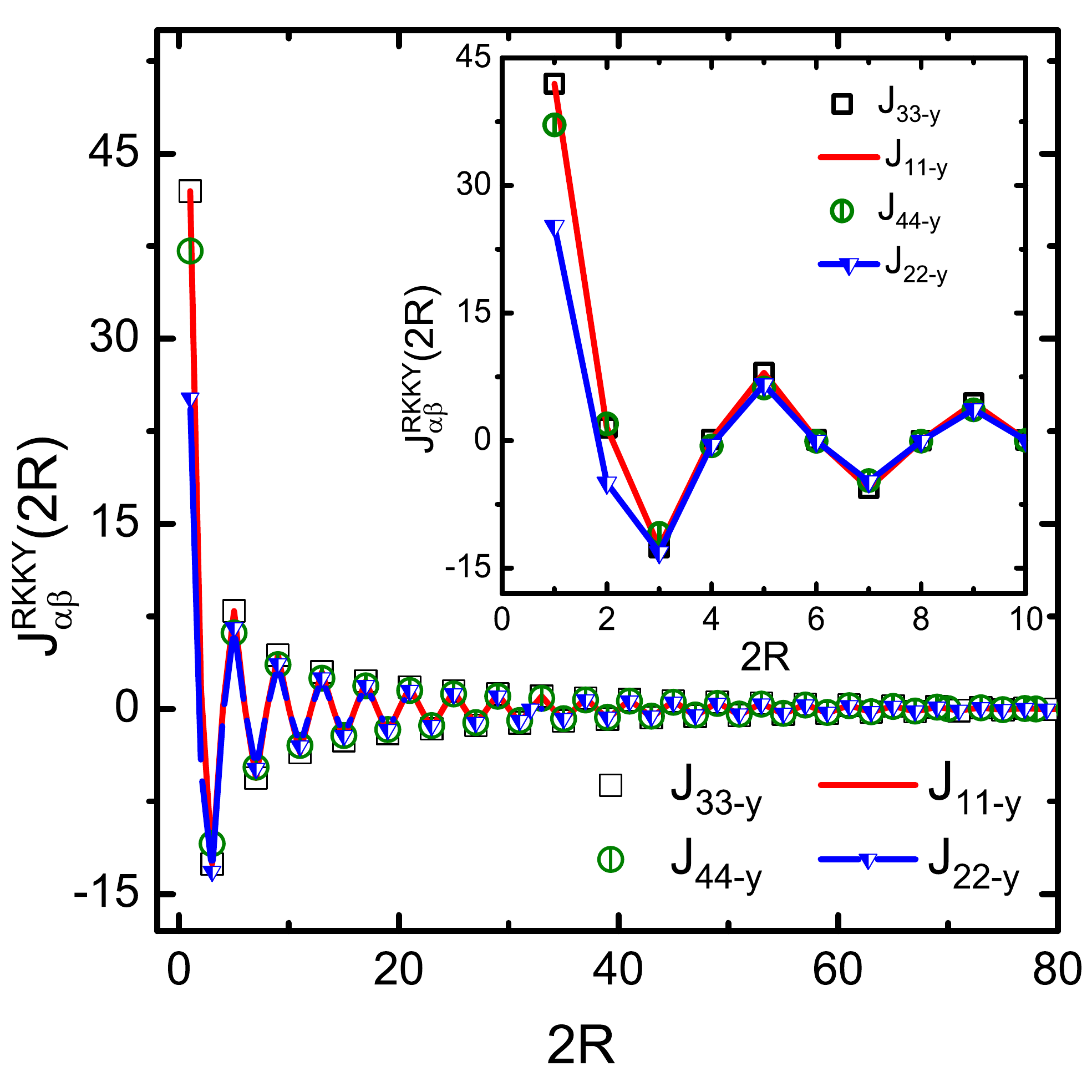}\label{fig:f5b}}
  \hfill  
  \subfloat[]{\includegraphics[width=0.248\textwidth]{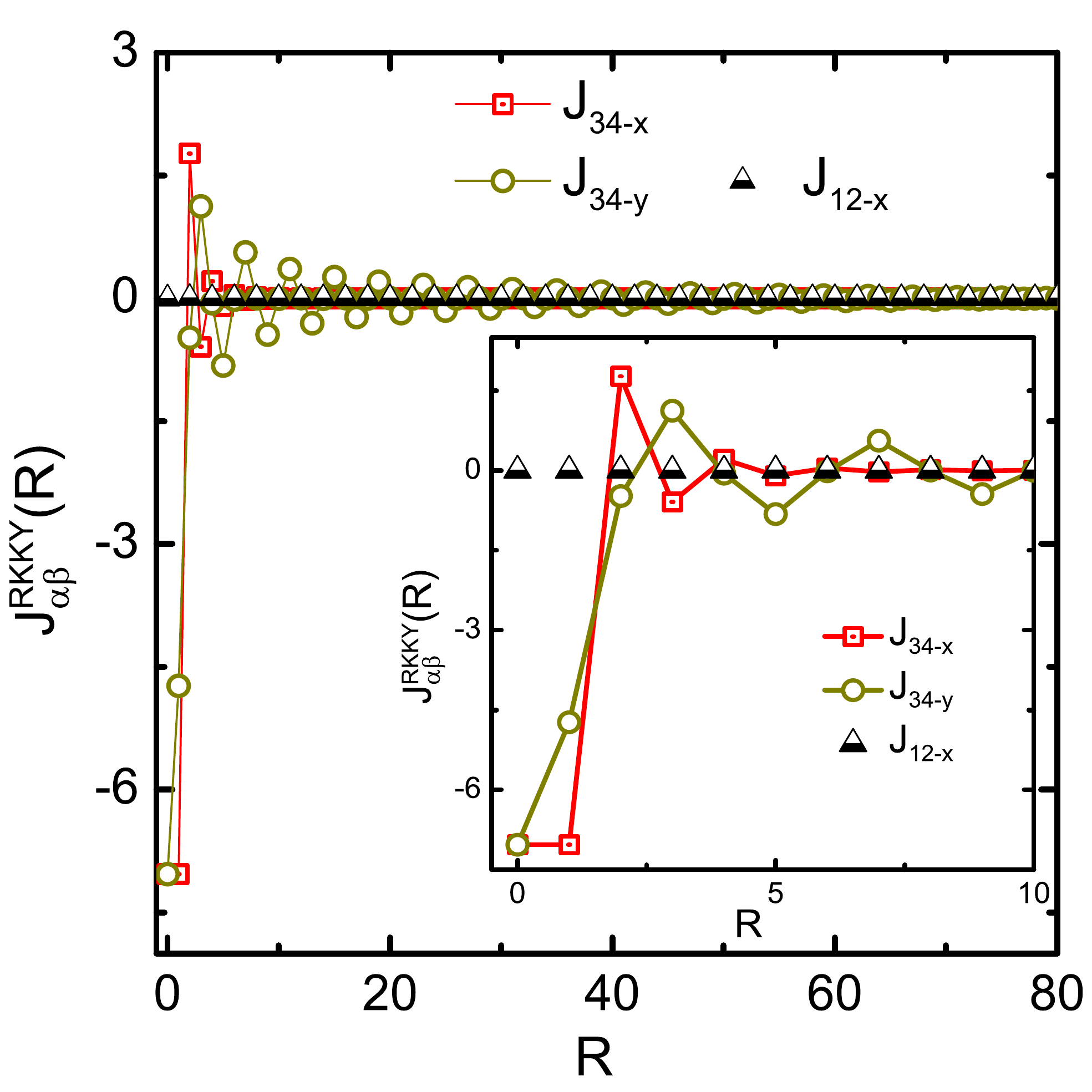}\label{fig:f5c}}
  \subfloat[]{\hspace{-0.11cm}\includegraphics[width=0.247\textwidth]{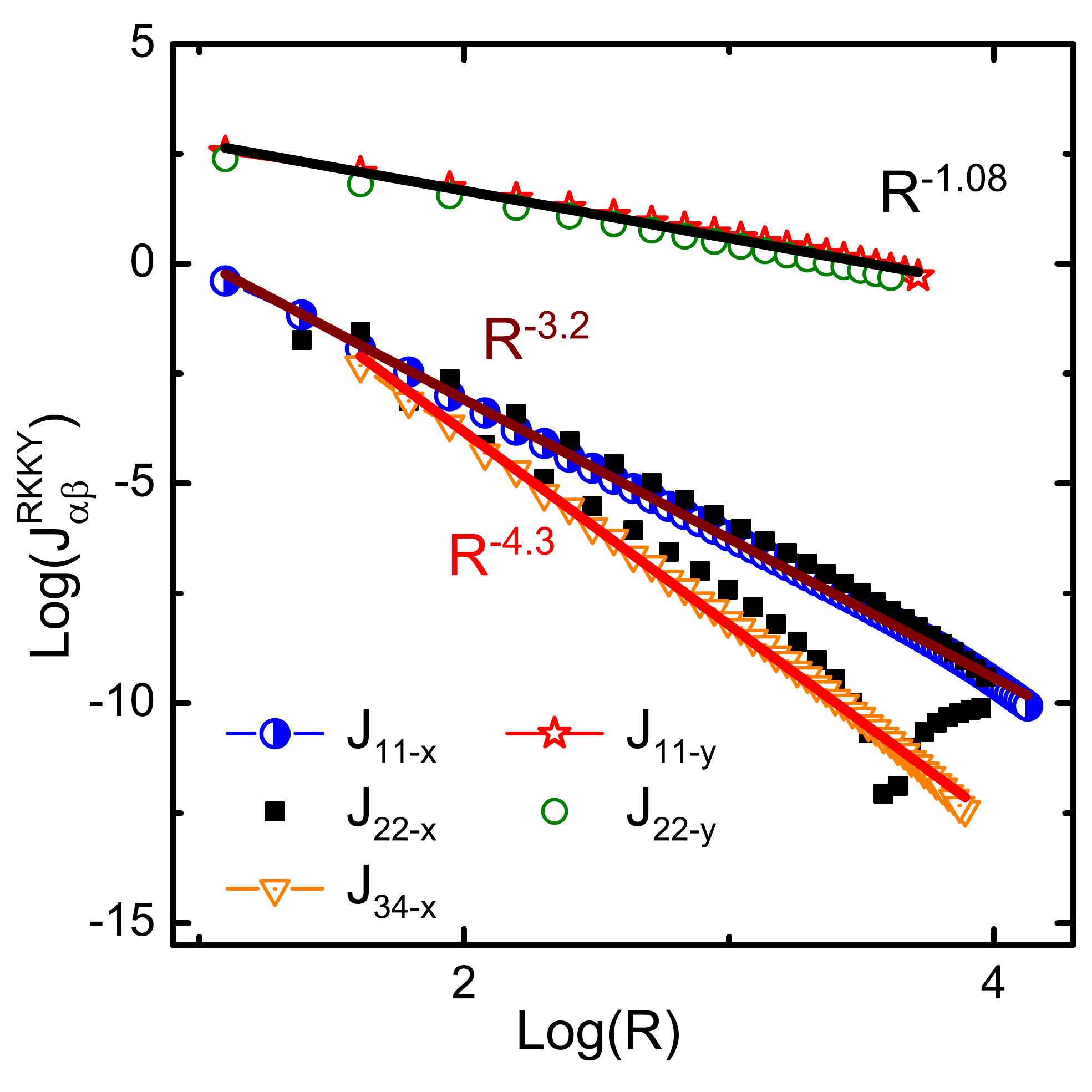} \label{fig:f5d}}
  \hspace*{\fill} 
  \caption{ \small Real-space RKKY interaction in $\pi$-flux lattice for same Ising spin pairs along the (a) x-direction and (b) y-direction  (c) the nearest-neighbor interactions of Ising spin pairs along the x- and y-directions. Both insets of figures zoom in the interacting within ten unit cells. (Here $\vec{R} = \vec{R}_j - \vec{R}_i$ is the distance vector of two unit cells $i$ and $j$, the distance between two unit cells along the  x- and y-directions are $R$ and $2R$, respectively). (d) The log-log plot of $J_{\alpha \beta}^\text{RKKY}$ via distance $R$ for the Ising spin pairs.}
  \label{fig5}
\end{figure}
	From the first-order and second-order perturbations in the momentum space, we perform their Fourier transformations into the real space. The first-order effective energies of 0- and $\pi$-flux models exist only when $\vec{q} = \vec{0}$ ($\delta_{\vec{q}, \vec{0}}$ terms in the equations (\ref{1-zero}) and (\ref{1-pi})). So, they are calculated exactly by taking integration over the Brillouin zone to provide a unique value depending linearly on the parameter $\xi$. The second-order energies are computed approximately according to the strength of Ising spin pair interactions.
	
	Figure~\ref{fig4} shows the real interacting pairs in 0-flux lattice such as Ising 1--1 (a pair of two Ising spins 1) and 1--2 couples (Figs.~\ref{fig:f4a} and \ref{fig:f4b}). Both interactions are decaying rapidly within one or two lattice distance $R$. Comparing with the normal RKKY interaction in the square lattice \citep{Allerdt}, our interacting functions are anisotropic and distinct along the x- and y-directions. For example, with the distance of one unit cell along the x-direction, the $J_{11}^\text{RKKY}(R = 1)$ value shows a negative sign with the factor $-(\xi/4\pi^2)^2$, that gives antiferromagnetic coupling between two Ising spins. However, along the y-direction, that coupling shows a positive sign that corresponds to FM coupling with three times larger magnitude (inset of Fig.~\ref{fig:f4a}). However, the interaction between different spins are antiferromangetically along both the x- and y-directions (Fig.~\ref{fig:f4b}). The long-tail interaction of Ising spins are considered carefully by taking $\log-\log$ plot in Fig.~\ref{fig:f4c}. Our results are consistent with the power decaying rate $R^{-1.85}$ that is closed to known value of RKKY interaction in two-dimensional lattice ($R^{-2}$) \citep{Allerdt}. 
	
	Because of fast decaying in the real-space interaction, we take the value of the closest distance with considering of different Ising spin coupling is the nearest-neighbor $J_{\text{NN}} = J_{\alpha \beta }^\text{RKKY} = 317.57(\xi/4\pi^2)^2$ with $\alpha \neq \beta$, and same Ising spins coupling is the next-nearest-neighbor $J_{\text{NNN}}$ along the x- and y-directions with $J_{\alpha \alpha - x  }^\text{RKKY} = 17.57(\xi/4\pi^2)^2$, and $J_{\alpha \alpha - y  }^\text{RKKY} = -60.3(\xi/4\pi^2)^2$, respectively. The magnitude of the nearest-neighbor coupling is much larger than the next-nearest one. The second-order effective energy for the 0-flux lattice is calculated by semi-analytic method:	
\begin{equation} \label{E_effective}
\begin{split}
E^{\text{RKKY}}_{0-\text{flux}} & \approx J_{\text{NN}} \sum_{ \langle \mu , \nu \rangle}  \sigma^{z}_{\mu} \sigma^{z}_{\nu} +  J_{\text{NNN}} \sum_{ \langle \zeta , \eta \rangle}  \sigma^{z}_{\zeta} \sigma^{z}_{\eta}.
\end{split}
\end{equation}
Where $\langle \mu , \nu \rangle$ and $\langle \zeta , \eta \rangle $ are the sum of all Ising spin pairs with distances of $\sqrt{2}/2$ and 1, respectively. With the coexistence of the nearest and next-nearest couplings between spins, the formation of magnetic order becomes frustrasted. That effect is similar to the observation of spin-nematic model in heavy fermion \chem{LiCuVO_4} compounds \citep{Orlova}. This compound is a typical example of competing effect in spin chain with the exisence of nearest-neighbor FM and next-nearest-neighbor AFM couplings. We need the exact diagonalization method to solve that problem and define the correct magnetic order.
  	    
 	In Fig.~\ref{fig5}, we calculate different coupling terms between four Ising spins of the $J_{\alpha \beta}^\text{RKKY}$ matrix (with $\alpha, \beta =  $1, 2, 3 and 4). Quite similar to the 0-flux case, $\pi$-flux results show two distinct interacting directions. Along the x-direction, the effective interactions is decaying rapidly within four to five lattice distance (Fig.~\ref{fig:f5a}). However, we observe the strong sign-changing oscillation along the y-direction even though the interacting distance between spins is raised in double (Fig.~\ref{fig:f5b}). That phenomenon is completely opposite to the no sign-changing oscillation of RKKY interaction in graphene  explained by collapsing of Fermi surface to Dirac points \citep{Kotov}. We believe that the sign-changing oscillation is also determined by the momentum function $ M_{\alpha \beta}(\vec{k}, \vec{q})$ outside of the Lindhard function. Interestingly, this model shows the magnitude of the nearest-neighbor are smaller than the next-nearest-neighbor one. For example, the nearest-neighbor coupling of Ising spin 1--2, 2--3 pairs are nearly vanished (Fig.~\ref{fig:f5c}) whereas the magnitude of Ising spin 3--4 are smaller $J_{34} =  6(\xi/2\pi^2)^2$ comparing with $J_{11-y} = 45(\xi/2\pi^2)^2$. Along the x-direction, interactions of Ising spin 1--1 and 3--3 couples have similar strength and ferromagnetic, but spin 2--2 and 4--4 couple ones are antiferromagnetic (inset of Fig.~\ref{fig:f5a}). Along the y-direction, both of them behave similarly as ferromagnetic interactions with larger value than x-direction ones (inset of Fig.~\ref{fig:f5b}). Because there are of ten couplings in two different directions, the magnetic order of the system is hard to predict by this method. 
 	
	Figure~\ref{fig:f5d} shows $\log-\log$ plots of $J_{\alpha \beta}^{\text{RKKY}}$ via distance $R$. We have a trouble to find the consistent decaying rate of that model. Along the y-direction, because of sign-changing oscillation, we take the maximum peaks of plot in Fig.~\ref{fig:f5b} to calculate $\log$ value, so they decay with power of $ R^{-1.08}$. The long-tail of nearest coupling 3--4 pair is the fastest decaying in the x-direction with $R^{-4.3}$ rate. Other x-direction couplings are decaying rate with power of $R^{-3}$. Our results show different behavior comparing with the long-distance limit of RKKY interaction in the graphene with decaying rate $R^{-3}$, even though we have similar form of interband Lindhard function and Dirac points. That issue may be interpreted by the effect of the momentum function $M_{\alpha \beta}(\vec{k}, \vec{q})$ outside of Lindhard function.    
\begin{figure*}[!ht] \label{fig6}
\centering
  \subfloat[]{%
    \includegraphics[width=.85\textwidth]{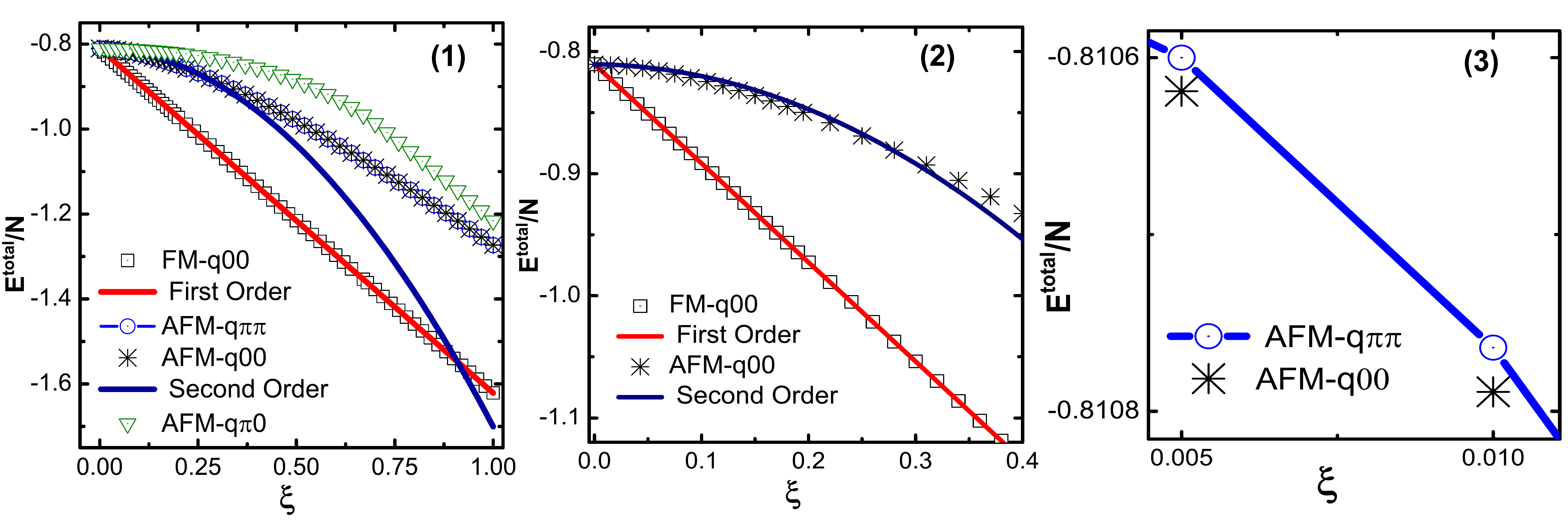} \label{fig:f6a}}
    \hfill 
   \subfloat[]{%
    \includegraphics[width=.85\textwidth]{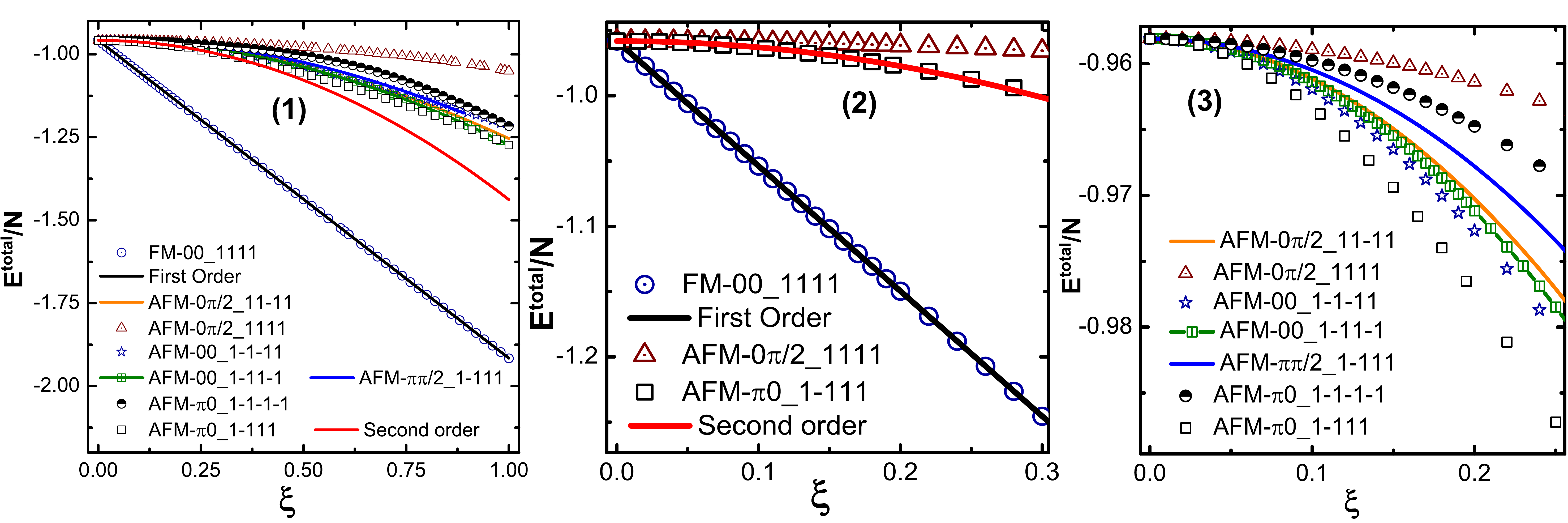} \label{fig:f6b}}
 \caption{\small The dependence of total energy $E^{\text{total}} $ on coupling parameter $\xi$ calculating from both semi-analytic (\textit{First Order} and \textit{Second Order} curves) and exact diagonalization methods with different magnetic orders for (a) the 0-flux lattice with FM phase (two spin-ups in one unit cell in the Fig.~\ref{fig:f1a}), and AFM orders (one spin-up and one spin-down in each unit cell), (\textit{AFM-q$\pi 0 $} is meant that we set up Ising spin 1 -- spin-up and Ising spin 2 -- spin-down in first unit cell, with the magnetic ordering vector $\vec{Q} = (\pi, 0)$) (b) FM order (with putting of 4 spin-ups in one unit cell in Fig.~\ref{fig:f1b}) and different AFM configurations in the $\pi$-flux lattice. (Here, the label of \textit{AFM-$\pi0$\_1-111} is meant that initial unit cell includes three spin-ups at the positions of Ising 1, 3 and 4, and one spin-down at the position of spin 2, (Fig.~\ref{fig:f1b}), and the magnetic ordering vector $\vec{Q} = (\pi, 0)$). }
\end{figure*}
\subsection{Energy configurations and magnetic orders}
\begin{figure}[!h]
\hspace*{\fill}
  \subfloat[]{\includegraphics[width=0.23\textwidth]{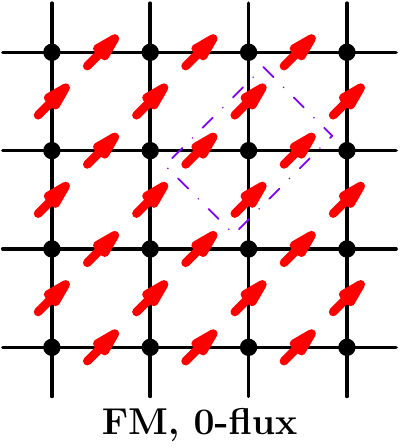}\label{fig:f7a}  }\hfill
  \subfloat[]{\includegraphics[width=0.23\textwidth]{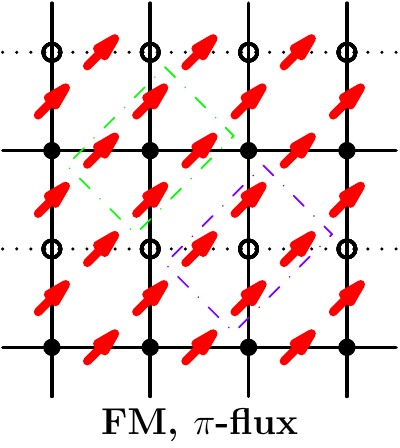}\label{fig:f7b}} 
  \\
  \subfloat[]{\includegraphics[width=0.23\textwidth]{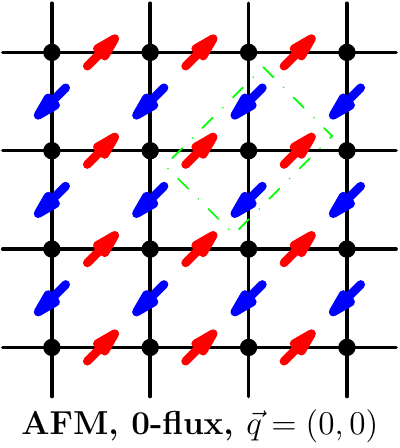}\label{fig:f7c} } \hfill
   \subfloat[]{\hspace{-0.2cm}\includegraphics[width=0.23\textwidth]{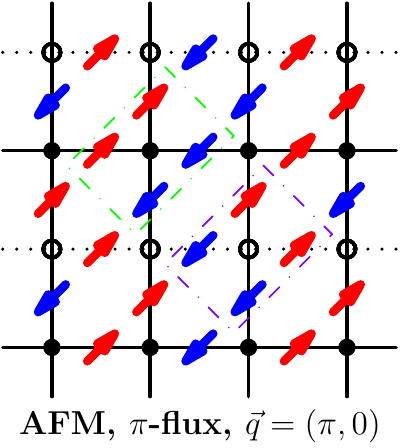}\label{fig:f7d}}
\hspace*{\fill}     
  \caption{\small The first-order ferromagnetic order of (a) the 0-flux model, (where black filled circle is the fermion and red arrow indicates Ising spin-up) and (b) the $\pi$-flux model, (here, black filled circle -- fermion A, and black open circle -- fermion B, and the red arrow -- Ising spin-up). Second-order antiferromagnetic orders of (c) the 0-flux and (d) $\pi$-flux models, (where blue arrow indicates Ising spin-down.) }
  \label{fig7}
\end{figure}
	We take advantage of the numerical calculation to search for the ground-state energy for each model. The total energy is calculated by exact diagonalization, and depended on the initial setup Ising spins at one unit cell, magnetic ordering vector $\vec{Q}$ and coupling parameter $\xi$. Figures \ref{fig:f6a} and \ref{fig:f6b} show different energy configurations of the 0- and $\pi$-flux lattices calculating both semi-analytic and exact diagonalization methods. We observe that the FM order is the lowest trivial energy (ground state) that comprises in both methods for two lattices. Our results show perfect matching between the semi-analytic method -- \textit{First Order} line and exact calculation -- \textit{FM-q00} open square in 0-flux model, or \textit{First Order} line and open circle \textit{FM-00-1111} in the $\pi$-flux (see Fig.~\ref{fig:f6a}{\small (2)} and Fig.~\ref{fig:f6b}{\small (2)} or $0\%$ error in Table.~\ref{table1}). It is meant that the non-interacting energy dispersion of the system is stretched out with $1+\xi$ amplitude because each hopping integral is increased linearly with factor $\xi$. If the ferromagnetic order is set up, the first-order perturbing energy is dominated. 
	
	We visualize the FM orders of two lattices in Fig.~\ref{fig:f7a} and Fig.~\ref{fig:f7b}. In the 0-flux lattice, there is a single fermion basis that surrounding by four Ising spin-ups (red arrow in the Fig.~\ref{fig:f7a}). Similarly, two different fermions in the $\pi$-flux couple four Ising spins-ups in four directions (dash violet and green boxes in the Fig.~\ref{fig:f7b}).

	The total energy of the system is generalized by:	       
\begin{equation}
E^{\text{total}} = E_0 + \xi A \sum_{\alpha} \sigma_{\alpha}^{z} + \xi^2 B \sum_{\alpha < \beta} \sigma_{\alpha}^{z} \sigma_{\beta}^{z},
\end{equation}	
where $A$ and $B$ are some constant coefficients that are extracted from the effective calculation in the semi-analytic method. $ \sum_{\alpha}$ is the sum of all Ising spins in lattices, and $\sum_{\alpha < \beta} $ is the sum of Ising pairs. So, $A\sum_{\alpha} \sigma_{\alpha}^{z} $ and $B\sum_{\alpha < \beta} \sigma_{\alpha}^{z} \sigma_{\beta}^{z} $ terms are exact first and second coefficients, respectively (listed in the Table \ref{table1}). For exact diagonalization method, they are interpolated from the plot of $E^{\text{total}}  -  \xi$. Detail calculations of $A$ and $B$ coefficients are found in the section 2 of Appendix~C. The $E^{\text{total}}$ includes the linear or quadratic forms of coupling parameter $\xi$.
	
	If our lattices are set up antiferromagnetically, the first-order perturbation energy is vanished ($\sum_{\alpha} \sigma_{\alpha}^{z} = 0$). For semi-analytic method, we follow the equation~(\ref{E_effective}) to calculate the effective energy of the 0-flux model. Unlike to unique FM magnetic order, several AFM configurations appear in the 0-flux and $\pi$-flux models.
	 
	In the Fig.~\ref{fig:f6a}{\small(1)}, we have seen that two AFM orders of 0-flux lattice: \textit{AFM-q00} (black star symbol) and \textit{AFM-q$\pi\pi$} (blue line and circle) behave similarly. However, when those plots are zoomed in the high enough resolution, the energy curve of  \textit{AFM-q00} order is lower value than \textit{AFM-q$\pi\pi$} one (Fig.~\ref{fig:f6a}{\small(3)} ). So, the advantage of exact diagonalization method over semi-analytic one is neglecting the singular point, and we have found the correct magnetic ordering vector $\vec{Q} = (0,0)$. Figure~\ref{fig:f6a}{\small(2)} shows agreement between semi-analytic and exact calculation results with $\xi \leq 0.2$ known as the fundamental property of perturbation theory -- weak coupling limit. The percent error of two methods is quite high as $12.7\%$ (Table \ref{table1}) because of the existence of singular point and approximation in the semi-analytic computation. 
	
	That AFM order with the magnetic ordering vector $ \vec{Q} = (0, 0)$ is visualized in the Fig.~\ref{fig:f7c}, and there are two spin-ups along the x-direction and two spin-downs along the y-direction. If we look at the $45^{\circ}$ rotation of the square lattice, that magnetic order is exact N\'{e}el state that is comprised with typical RKKY ground state in the square lattice \citep{Budapest}. The effect of spin-up enhances hopping (or weak bond), and the spin-down retards the hopping (or strong bond) of fermion from one site to the other. If the lattice is applied the electric field, the electric current will flow anisotropic, and be higher conductivity in the x-direction than y-direction. 
	
	In Fig.~\ref{fig:f6b}, we plot the energy $ E^{\text{total}}$ versus coupling parameter $\xi$ curves of the AFM configurations in $\pi$-flux lattice. Because of four different Ising spins at one unit cell, we have investigated more spin configurations than the 0-flux one. Figure \ref{fig:f6b}{\small(2)} zooms in the AFM energy curves calculated from \textit{Second order} plot by semi-analytic method (red solid line) and \textit{AFM$\pi 0 \_$1-111} plot by exact diagonalization (open square symbol) with setting up of three spin-ups positioned at Ising 1, 3, and 4, one spin-down for Ising 2, and the magnetic ordering vector $ \vec{Q} = (\pi, 0)$. Two energetic ($E^{\text{total}}-\xi$) plots of the methods match very well with $1.0\%$ error of the different coefficients (Table \ref{table1}). We believe thi magnetic order is the ground-state energy of AFM configurations in the Fig.~\ref{fig:f6b}{\small(3)}. That order preserves the $Z_2$ symmetry when I reversed all signs of Ising spins. Analogous to the 0-flux lattice, the semi-analytic result is only fitted with exact diagonalization at the limit of $\xi  \leq 0.3$. 

	The AFM order is visualized in Fig.~\ref{fig:f7d} with two distinct fermions (in the dash violet and green boxes). Each fermion is coupled with four surrounding Ising spins. Fermion A (filled black circle) first coupled with two spin-ups in the East and South directions, and two spin-downs in the West and North directions (green box), so the next-nearest fermion A along the x-direction reverses all signs of Ising spins. Fermion B has a similar behavior (open circle in the violet box). However, along the y-direction, the the sign of Ising spin are unchanged. That follows the magnetic ordering order $\vec{Q} = (\pi, 0)$, or it is like a spin wave along the x-direction. Therefore, the AFM state of $\pi$-flux is different from 0-flux one.
\begin{table}
   \caption{ Summarizing coefficients of the effective energy are calculated in both semi-analytic and exact diagonalization methods of the 0- and $\pi$-flux models: \label{table1}}
   \centering	
  \begin{tabular}{|C{1.5cm}|C{1.5cm}|C{1.5cm}|C{1.5cm}|C{1.5cm}|}
    \hline
    Models 
    &    
      \multicolumn{2}{|c|}{0-flux} &
      \multicolumn{2}{|c|}{$\pi$-flux}
      \\
      \hline
    Methods & $1^{\text{st}}$ order & $2^{\text{nd}}$ order & $1^{\text{st}}$ order& $2^{\text{nd}}$ order \\
    \hline
    Semi & $-0.811$ & $-0.890$ & $-0.958$ & $-0.48$  \\
    \hline
    Exact & $-0.811$ & $-1.02$ & $-0.958$ & $-0.484$ \\
    \hline
    \% error & 0.0\% & 12.7\% & 0.0\% & 1.0\% \\
    \hline
  \end{tabular}
\end{table}
\section{Conclusions}
	We consider the model of Ising spins on the links of the square lattices, coupled to the fermion charge fluctuations for the 0-flux and $\pi$-flux cases. The unit cell of the $\pi$-flux lattice doubles of 0-flux one. The Brillouin zone changed from symmetrical square of 0-flux to asymmetrical rectangular of $\pi$-flux lattice. At the half-filling, the tight binding parts of Hamiltonian provide continuous metallic band and semimetallic Dirac bands for the 0- and $\pi$-flux lattices, respectively. We observe the Fermi surface with the nesting vectors, and Fermi points in Dirac bands for two cases.
	
	Use the semi-analytic perturbation theory and exact diagonalization method, we achieve different magnetic orders and ground state energies based on spin configurations. With the first-order perturbation, both kinds of models show the trivial FM ground state, and ground state energy that is linear dependent of coupling parameter $\xi$. That is due to spin-up enhances the hopping magnitude with the factor of $1 + \xi$. We can use our models or depleted Anderson's model \citep{Tivinidze} to construct the FM magnetic order in the square lattice.
	 
	 When the Ising spin configuration is set up antiferromagnetically, the second-order perturbation is considered. The ground state energy now is following the quadratic function of coupling parameter $\xi$. The strong agreement between the semi-analytic and exact diagonalization methods only exists with the coupling parameter $\xi \ll 1$. We finally construct lowest energy AFM configuration with the N\'{e}el state of the $45^{\circ}$ rotation $\vec{Q} = (0,0)$, and spin-wave AFM oscillating along the x-direction with the ordering vector $\vec{Q} = (\pi,0)$ for the 0-flux and $\pi$-flux models, respectively.    
	         
	 Our achieving results contribute to the rich variety of phenomena in the Ising-nematic square lattices \citep{Xu, Assaad2, Gazit}. Particularly, the FM state exists in the 2D experimental observation can be explained by our models \citep{Huang}. Moreover, we consider a case of coupling $\xi$ between spin and fermion in the weak limit. Our future work will continue with inserting the other interaction like Hubbard term \citep{Otsuka, Assaad1}, and tuning the filling factor \citep{Gazit} to those lattices. 
\section{Acknowledgement}
	Numerical calculations have been carried out on machines hosted at the Mississippi Center for Supercomputing Research, University of Mississippi.
\appendix
\section{Explicit real-space Hamiltonian }
\subsection{Zero-flux model}
	The Hamiltonian of 0-flux lattice is written explicitly in real space with the lattice translation vector $\vec{R} = m\vec{a}_1 + n\vec{a}_2$ ($m$, $n$ are integers):	
\begin{equation}  \label{H0}
\centering
 \hat{H}_0 = -t \sum_{\vec{R}}[(c_{\vec{R}+\vec{a_1}}^\dagger c_{\vec{R}} + c_{\vec{R}}^{\dagger} c_{\vec{R}+\vec{a_1}} ) + (c_{\vec{R}+\vec{a_2}}^\dagger c_{\vec{R}} + c_{\vec{R}}^{\dagger} c_{\vec{R}+\vec{a_2}}) ],   
\end{equation}   
\begin{equation} \label{H1}
\centering
\hat{H}_1 = -\xi\sum_{\vec{R}}[\sigma^z_{\vec{R}, 1}(c_{\vec{R}+\vec{a_1}}^\dagger c_{\vec{R}} + c_{\vec{R}}^{\dagger} c_{\vec{R}+\vec{a_1}} ) + \sigma^z_{\vec{R}, 2}(c_{\vec{R}+\vec{a_2}}^\dagger c_{\vec{R}} + c_{\vec{R}}^{\dagger} c_{\vec{R}+\vec{a_2}}) ].      
\end{equation} 
\subsection{Discrete Fourier transformation identities}
The discrete Fourier transformation for fermion $c_{\vec{R}}^\dagger $ and Ising spin $\sigma_{\vec{R},\alpha}^z$ from the real space to momentum space are:
\begin{align}
c_{\vec{R}}^\dagger  & = \frac{1}{\sqrt{N}}\sum_{\vec{k}}e^{-i\vec{k}.\vec{R}}c_{\vec{k}}^\dagger \quad (N: \text{number of unit cells} ),\\
\sigma_{\vec{R},\alpha}^z & = \frac{1}{N}\sum_{\vec{q}}e^{-i\vec{q}.\vec{R}}\sigma_{\vec{q},\alpha}^z          \quad (\text{with} \quad \alpha = 1, 2).
\end{align}
\subsection{$\pi$-flux model}
	The general Hamiltonian of the $\pi$-flux lattice is the form of vector $\vec{R}$ in the real space:
 \begin{widetext}
\begin{eqnarray} 
   \hat{H}  & = &  \sum_{\vec{R}} \Big [  - (t + \xi  \sigma_{\vec{R}, 1}^z)c_{\vec{R}+{\vec{a}_1},\text{A}}^\dagger  c_{\vec{R}, \text{A}}    -   (t + \xi \sigma_{\vec{R}-{\vec{a}_1}, 1}^z) c_{\vec{R}-{\vec{a}_1},\text{A}}^\dagger c_{\vec{R},\text{A}}     
     -  (t + \xi \sigma_{\vec{R},2}^z)(c_{\vec{R},\text{B}}^\dagger c_{\vec{R},\text{A}} + c_{\vec{R},\text{A}}^\dagger c_{\vec{R},\text{B}})  \nonumber 
     \\
    && + \>  (t + \xi \sigma_{\vec{R},3}^z)c_{\vec{R}+{\vec{a}_1},\text{B}}^\dagger c_{\vec{R},\text{B}}  
                + (t + \xi \sigma_{\vec{R}-{\vec{a}_1},3}^z) c_{\vec{R}-{\vec{a}_1}, \text{B}}^\dagger c_{\vec{R}, \text{B}}  -  (t + \xi \sigma_{\vec{R}-{\vec{a}_2}, 4}^z)c_{\vec{R}-{\vec{a}_2}, \text{B}}^\dagger c_{\vec{R}, \text{A}}  -(t + \xi \sigma_{\vec{R},4}^z)c_{\vec{R}+{\vec{a}_2},\text{A}}^\dagger c_{\vec{R},\text{B}} \Big ].                                                         
\end{eqnarray}
	We separate into two parts: tight-binding Hamiltonian $\hat{H}_0$ and interacting Hamiltonian $\hat{H}_1$:	
\begin{equation}
 \hat{H_0}  =  -t\sum_{\vec{R}} \big [ (c_{\vec{R}+{\vec{a}_1}, \text{A}}^\dagger c_{\vec{R}, \text{A}} +  c_{\vec{R}-{\vec{a}_1}, \text{A}}^\dagger c_{\vec{R}, \text{A}}) 
   + (c_{\vec{R}, \text{A}}^\dagger c_{\vec{R}, \text{B}}   + c_{\vec{R}, \text{B}}^\dagger c_{\vec{R}, \text{A}}) 
   - (c_{\vec{R} + \vec{a}_1, \text{B}}^\dagger c_{\vec{R}, \text{B}} + c_{\vec{R} - \vec{a}_1, \text{B}}^\dagger c_{\vec{R}, \text{B}})  + 
      (c_{\vec{R}-{\vec{a}_2}, \text{B}}^\dagger c_{\vec{R}, \text{A}} + c_{\vec{R}+{\vec{a}_2}, \text{A}}^\dagger c_{\vec{R}, \text{B}}) \big ],    
\end{equation} 
\begin{eqnarray}
\hat{H}_{1} & = & -\xi\sum_{\vec{R}}\big [ \sigma_{\vec{R},2}^zc_{\vec{R}+{\vec{a}_2}, \text{A}}^\dagger c_{\vec{R}, \text{B}} + \sigma_{\vec{R}-{\vec{a}_2},2}^z c_{\vec{R}-{\vec{a}_2}, \text{B}}^\dagger c_{\vec{R}, \text{A}}    
                  +  \sigma_{\vec{R},1}^z  c_{\vec{R}+{\vec{a}_1}, \text{B}}^\dagger c_{\vec{R}, \text{B}} + \sigma_{\vec{R}-{\vec{a}_1},1}^z c_{\vec{R}-{\vec{a}_1}, \text{B}}^\dagger c_{\vec{R}, \text{B}}                  
                   +  \sigma_{\vec{R},4}^z (c_{\vec{R}, \text{B}}^\dagger c_{\vec{R}, \text{A}} + c_{\vec{R}, \text{A}}^\dagger c_{\vec{R}, \text{B}})  \nonumber
                    \\
                    && - \> \sigma_{\vec{R},3}^z c_{\vec{R}+{\vec{a}_1}, \text{A}}^\dagger c_{\vec{R}, \text{A}} - \sigma_{\vec{R}-{\vec{a}_1},3}^z c_{\vec{R}-{\vec{a}_1}, \text{A}}^\dagger c_{\vec{R}, \text{A}} \big ].     
\end{eqnarray}
The real-space Hamiltonians are then transformed into to the momentum space using the discrete Fourier transformations above. After that, we use the Fermi-sea ground state to calculate the first- and second-order effective interactions and energies. (Detail calculation is described in my master thesis \citep{HDo}, or in the reference \citep{Budapest}).

\section{Function definitions}
\subsection{Second-order RKKY interaction in the 0-flux lattice}
All terms of 2$\times$2 $ J_{\alpha \beta}(\vec{k}, \vec{q})$ matrix are listed below:
\begin{equation}
\left.\begin{aligned}
J_{11} (\vec{k}, \vec{q})  & =  2 + e^{i(2k_x + q_x)} + e^{-i(2k_x + q_x)}  = 4\cos^2(k_x + \frac{q_x}{2}),
\\
J_{22} (\vec{k}, \vec{q})  & =  2 + e^{i(2k_y + q_y)} + e^{-i(2k_y + q_y)} = 4\cos^2(k_y + \frac{q_y}{2}), 
\\
J_{12}(\vec{k}, \vec{q})  & =  e^{-i(k_x + k_y + q_y)} + e^{i(k_x + k_y + q_x)} + e^{-i(k_x - k_y)} 
 +  e^{i(k_x - k_y + q_x- q_y)}, 
 \\
J_{21}(\vec{k}, \vec{q}) & =  e^{i(k_x + k_y + q_y)} + e^{-i(k_x + k_y + q_x)} + e^{i(k_x - k_y)} 
 +  e^{-i(k_x - k_y + q_x- q_y)}.
\end{aligned}\right.
\end{equation}
\subsection{Second-order RKKY interaction in the $\pi$-flux lattice}
There are some useful functions:
\begin{align}
u{(\vec{k})}   & = \sqrt{\cos^2{k_x} + \cos^2{k_y}} -\cos{k_x},  \\
v({\vec{k}})  & = \sqrt{2(\cos^2{k_x} + \cos^2{k_y} - \cos{k_x}\sqrt{\cos^2{k_x} + \cos^2{k_y}})},  \\
u{(\vec{k}, \vec{q})}  & = \sqrt{\cos^2(k_x + q_x) + \cos^2(k_y + q_y)} -\cos(k_x + q_x),  \\
v({\vec{k}, \vec{q}})  & = \sqrt{2[\cos^2(k_x + q_x) + \cos^2(k_y+q_y) - \cos(k_x + q_x)\sqrt{\cos^2(k_x+q_x) + \cos^2(k_y+q_y)} ]}. 
\end{align}
All terms of 4$\times$4 $ M_{\alpha \beta}(\vec{k}, \vec{q})$ matrix are listed below:
\begin{eqnarray}
\left.\begin{aligned}
M_{11}(\vec{k}, \vec{q}) & = & 4\cos^2(k_x + \frac{q_x}{2}) \frac{ \cos^2(k_y) \cos^2(k_y + q_y) }{v^2(\vec{k}) v^2(\vec{k} + \vec{q})},
\end{aligned}\right. 
\end{eqnarray}
\begin{equation}
%
M_{33} (\vec{k}, \vec{q})      =  \big [ 2 + 2 \cos(2k_x + q_x) \big ] \frac{ u^2(\vec{k}) u^2(\vec{k}+\vec{q})}{v^2(\vec{k}) v^2(\vec{k}+\vec{q}},
\end{equation}
%
\begin{eqnarray}
M_{13}(\vec{k}, \vec{q})  & = & \frac{ [1 + \cos(2k_x + q_x)] u(\vec{k}) u(\vec{k} + \vec{q}) [1 + e^{i2(k_y + q_y)}][1 + e^{i2k_y}]}{2v^2(\vec{k}) v^2(\vec{k} + \vec{q})}, 
\end{eqnarray}
\begin{eqnarray}
M_{31}(\vec{k}, \vec{q})  & = & \frac{ [1 + \cos(2k_x + q_x)] u(\vec{k}) u(\vec{k} + \vec{q}) [1 + e^{-i2(k_y + q_y)}][1 + e^{-i2k_y}]}{2v^2(\vec{k}) v^2(\vec{k} + \vec{q})}, 
\end{eqnarray}
%
\begin{eqnarray}
M_{14}(\vec{k}, \vec{q})  & = & \frac{ e^{i (k_x + q_x)} + e^{-ik_x} }{4 v^2(\vec{k}) v^2(\vec{k} + \vec{q})} \{ - u(\vec{k} + \vec{q}) [e^{i2k_y} + e^{i(4k_y + 2q_y)}] (1 + \cos{2k_y})  \nonumber
\\
&&  + \> u(\vec{k}) [1 + \cos(2k_y + 2q_y)][e^{-i2 q_y} + e^{-i2(k_y + q_y)}]  \}, 
\end{eqnarray}
%
\begin{eqnarray}
M_{41}(\vec{k}, \vec{q})  & = & \frac{ e^{-i (k_x + q_x)} + e^{ik_x} }{4 v^2(\vec{k}) v^2(\vec{k} + \vec{q})} \{ -u(\vec{k} + \vec{q}) [e^{-i2k_y} + e^{-i(4k_y + 2q_y)}] (1 + \cos{2k_y})  \nonumber
\\
&&  + \> u(\vec{k}) [1 + \cos(2k_y + 2q_y)][e^{i2 q_y} + e^{i2(k_y + q_y)}]  \},  
\end{eqnarray}
%
\begin{eqnarray}
M_{23}(\vec{k}, \vec{q})   =   \{ u(\vec{k}) [ 1 + e^{i2(k_y + q_y)}  ] - u(\vec{k} + \vec{q}) (1 + e^{i2k_y})  \}  
 \frac{ u(\vec{k}) u(\vec{k} + \vec{q})[e^{ik_x} + e^{-i(k_x + q_x)}]}{2 v^2(\vec{k}) v^2(\vec{k} + \vec{q})}, 
\end{eqnarray}
%
\begin{eqnarray}
M_{32}(\vec{k}, \vec{q})   =   \{ u(\vec{k}) [ 1 + e^{-i2(k_y + q_y)}  ] - u(\vec{k} + \vec{q}) (1 + e^{-i2k_y})  \} 
\frac{  u(\vec{k}) u(\vec{k} + \vec{q})[e^{-ik_x} + e^{i(k_x + q_x)}]}{2 v^2(\vec{k}) v^2(\vec{k} + \vec{q})}, 
\end{eqnarray}
%
\begin{eqnarray}  
M_{34}(\vec{k}, \vec{q})  & = & \frac{ u(\vec{k}) u(\vec{k} + \vec{q})[e^{-ik_x} + e^{i(k_x + q_x) ]} }{2v^2(\vec{k}) v^2(\vec{k} + \vec{q})} \big \{ u(\vec{k}) [ e^{-i2(k_y + q_y)} + e^{-i4(k_y + q_y)} ] 
 -  u(\vec{k} + \vec{q}) (1 + e^{i2k_y})  \big \},  
\end{eqnarray}
%
\begin{eqnarray}  
M_{43}(\vec{k}, \vec{q})  & = & \frac{ u(\vec{k}) u(\vec{k} + \vec{q})[e^{ik_x} + e^{-i(k_x + q_x) ]} }{2 v^2(\vec{k}) v^2(\vec{k} + \vec{q})} \big \{ u(\vec{k}) [ e^{i2(k_y + q_y)} + e^{i4(k_y + q_y)} ] 
 -  u(\vec{k} + \vec{q}) (1 + e^{-i2k_y})  \big \},  
\end{eqnarray}
\begin{eqnarray}
M_{22}(\vec{k}, \vec{q})  & = &  \frac{1}{2 v^2(\vec{k}) v^2(\vec{k} + \vec{q})} \big \{ - \big [ 1 +  \cos(2k_y) + \cos(2k_y + 2q_y) + \cos(2q_y) \big ]  u(\vec{k}) u(\vec{k} + \vec{q})     \nonumber 
\\
&&  + \> [1 + \cos(2k_y)] u^2(\vec{k} +\vec{q}) + [1 + \cos(2k_y + 2q_y)] u^2(\vec{k}) \big \},  
\end{eqnarray}
%
%
\begin{eqnarray}
M_{12}(\vec{k}, \vec{q})  & = &\frac{ [e^{-i k_x} + e^{i (k_x +q_x)}] }{4 v^2(\vec{k}) v^2(\vec{k} + \vec{q})} \big \{  u(\vec{k}) (1 +  e^{i2k_y}) [1 + \cos(2k_y + 2q_y)]       
  -  u(\vec{k}+\vec{q}) [1 + e^{i(2k_y + 2q_y) }] (1 + \cos{2k_y}) \big \}, 
\end{eqnarray}
\begin{eqnarray}
M_{21}(\vec{k}, \vec{q})  & = &\frac{ [e^{i k_x} + e^{-i (k_x +q_x)}] }{4 v^2(\vec{k}) v^2(\vec{k} + \vec{q})} \big \{  u(\vec{k}) (1 +  e^{-i2k_y}) [1 + \cos(2k_y + 2q_y)]      
  -  u(\vec{k}+\vec{q}) [1 + e^{-i(2k_y + 2q_y) }] (1 + \cos{2k_y}) \big \}, 
\end{eqnarray}
\begin{eqnarray}
M_{24}(\vec{k}, \vec{q})  & = &  \frac{1}{4 v^2(\vec{k}) v^2(\vec{k} + \vec{q})} \big \{ e^{i2k_y} (-1  - e^{i2(k_y + q_y)}) (1 + e^{-i2k_y}) u(\vec{k}) u(\vec{k} + \vec{q})
 +  e^{i2k_y}[2 + 2\cos(2k_y)] u^2(\vec{k} +\vec{q})  \nonumber 
\\
&& + \> e^{-i2(k_y + q_y)}[2 + 2\cos(2k_y + 2q_y)] u^2(\vec{k})   
 -  e^{-i2(k_y + q_y)} (1 + e^{i2k_y}) [1 + e^{-i2(k_y + q_y)}] u(\vec{k}) u(\vec{k} + \vec{q})   \big \},
\end{eqnarray}
%
\begin{eqnarray}
M_{42}(\vec{k}, \vec{q})  & = &  \frac{1}{4 v^2(\vec{k}) v^2(\vec{k} + \vec{q})} \big \{ e^{i2(k_y+q_y)} (1 + e^{-i2k_y}) (-1  - e^{i2(k_y + q_y)}) u(\vec{k}) u(\vec{k} + \vec{q}) 
 +  e^{-i2k_y}[2 + 2\cos(2k_y)] u^2(\vec{k} +\vec{q}) \nonumber
\\
&& + \> e^{i2(k_y + q_y)}[2 + 2\cos(2k_y + 2q_y)] u^2(\vec{k})   
 -  e^{-i2(k_y)} (1 + e^{i2k_y}) [1 + e^{-i2(k_y + q_y)}] u(\vec{k}) u(\vec{k} + \vec{q})   \big \},
\end{eqnarray}
%
\begin{eqnarray}
M_{44}(\vec{k}, \vec{q})  & = &  \frac{1}{2 v^2(\vec{k}) v^2(\vec{k} + \vec{q})} \big \{ - [ \cos(4k_y + 2q_y) + \cos(6k_y + 3q_y) + \cos(2k_y + 2q_y)    
 +  \cos(4k_y + 4q_y) ] u(\vec{k}) u(\vec{k} + \vec{q}) \nonumber 
 \\
 && + \> [1 + \cos(2k_y)] u^2(\vec{k} +\vec{q})  +   [1 + \cos(2k_y + 2q_y)] u^2(\vec{k}). 
  \big \}
\end{eqnarray}
\section{Example calculation}
\subsection{Derivation of static function $J_{11}(\vec{q}, \omega_n=0)$}
If we consider the $\vec{q}$ is very small, we can expand the Fermi-Dirac function $n_F(E_{\vec{k} + \vec{q}})$ and energy dispersion relation $E(\vec{k}+\vec{q}) $ of the Lindhard function in terms of $\vec{q} $. So the momentum-dependence of two spin 1 interaction is written: 
\begin{eqnarray}
J_{11}(\vec{q} \to \vec{0}) & = & \frac{1}{4\pi^2} \int_{-\pi}^{\pi} dk_x \int_{-\pi}^{\pi} dk_y -n^{\prime}(E_{\vec{k}}) 2 (1 + \cos{2k_x}) \quad
 =  \frac{1}{4\pi^2} \int_{-\pi}^{\pi} dk_x \int_{-\pi}^{\pi} dk_y \delta[\cos{k_x} + \cos{k_y}] 2 \cos^2{k_x} \nonumber
\\
				& = & \frac{4}{4\pi^2} \int_{-\pi}^{0} dk_x \int_{-\pi}^{\pi} dk_y \delta(\pi + k_x - k_y)\frac{\cos^2{k_x}}{\lvert\sin{k_y}\rvert} \quad
				 =  	\frac{2}{\pi}	\int_{-\pi}^{0} dk_x \frac{\cos^2{k_x}}{\sin{k_x}}.		
\end{eqnarray} 
\subsection{Calculation of second-order effective energy using semi-analytical methods}
	For 0-flux lattice, the real-space interactions calculating by semi-analytic method include the nearest-neighbor and next-nearest-neighbor pairs. When the lattice size of the system is $L$, the total number of Ising spins living at the bonds is $N_{\text{spin}} = 2L^2 $. We have $N$ number of Ising 1 and $N$ number of Ising spin 2. There are $2N$ numbers of antiferromagnetic Ising 1--2 and 2--1 pairs, and $N/2$ number of Ising 1--1 along the x- and y- directions.
\begin{equation}
E_{\text{Second Order}} = -[315.57 \times 4N + (60.3-15.57)\times 2N]\frac{\xi^2}{(4\pi^2)^2} = -0.89\times \xi^2.
\end{equation}  
So, the second-order coefficient in the 0-flux lattice is $B\sum_{\alpha < \beta} \sigma_{\alpha}^{z} \sigma_{\beta}^{z} = -0.89 $. The coefficients of the $\pi$-flux are calculated in the similar way.

	For exact diagonalization, when we obtain the total energy data, we plot them via the coupling parameter $\xi$ for different Ising spin configurations. Using power fitting, we find the first- and second-order coefficients.
\end{widetext}

\label{Bibliography}
\bibliographystyle{apsrev4-1}  
\bibliography{Bibliography}  
\end{document}